\documentstyle[10pt,epsfig,dp_delphititle,float,hangcaption,xspace,amssymb,
amsfonts,amsmath,amsthm,cite,graphicx]{dp_delphi}
\makeindex
\def\DpPaperGroup{EP-PH}
\def\DpPaperRef{2004-072}
\def\DpDate{10 December 2004}
\def\DpAuthors{DELPHI Collaboration}
\def\DpSubmit{(Accepted by Euro. Phys. Jour.)}
\def\DpTitle{{Bose-Einstein Correlations in \\W$^{\bf +}$ W$^{\bf -}$ events at LEP2}}

\newcommand{\braket}[1]{\langle #1 \rangle}

\newcommand{\p}{$\pm$\xspace}
\newcommand{\degr}{$^\circ$\xspace}

\begin{document}
\makeatletter
\newcount\@tempcntc
\def\@citex[#1]#2{\if@filesw\immediate\write\@auxout{\string\citation{#2}}\fi
  \@tempcnta\z@\@tempcntb\m@ne\def\@citea{}\@cite{\@for\@citeb:=#2\do
    {\@ifundefined
       {b@\@citeb}{\@citeo\@tempcntb\m@ne\@citea\def\@citea{,}{\bf ?}\@warning
       {Citation `\@citeb' on page \thepage \space undefined}}%
    {\setbox\z@\hbox{\global\@tempcntc0\csname b@\@citeb\endcsname\relax}%
     \ifnum\@tempcntc=\z@ \@citeo\@tempcntb\m@ne
       \@citea\def\@citea{,}\hbox{\csname b@\@citeb\endcsname}%
     \else
      \advance\@tempcntb\@ne
      \ifnum\@tempcntb=\@tempcntc
      \else\advance\@tempcntb\m@ne\@citeo
      \@tempcnta\@tempcntc\@tempcntb\@tempcntc\fi\fi}}\@citeo}{#1}}
\def\@citeo{\ifnum\@tempcnta>\@tempcntb\else\@citea\def\@citea{,}%
  \ifnum\@tempcnta=\@tempcntb\the\@tempcnta\else
   {\advance\@tempcnta\@ne\ifnum\@tempcnta=\@tempcntb \else \def\@citea{--}\fi
    \advance\@tempcnta\m@ne\the\@tempcnta\@citea\the\@tempcntb}\fi\fi}
 
\makeatother
\begin{titlepage}
\pagenumbering{roman}
\CERNpreprint{\DpPaperGroup}{\DpPaperRef} 
\date{{\small\DpDate}} 
\title{\DpTitle} 
\address{\DpAuthors} 
\begin{shortabs} 
\noindent
Bose-Einstein correlations (BEC) between final state particles in the reaction  $\rm{e^+e^- \to W^+W^- \to q_1\overline{q_2}q_3\overline{q_4}}$\hspace{2mm} have
been studied.
 Data corresponding to a total integrated luminosity of 550 pb$^{-1}$, recorded by the
DELPHI detector at centre-of-mass energies ranging from 189 to 209 GeV, were
analysed.
An indication for inter-W BEC between like-sign particles has been found at
the level of 2.4 standard deviations of the combined statistical and 
systematic uncertainties.

\vskip 2cm
\begin{center}
{\it
This paper is dedicated to the late Frans Verbeure.
Frans was a very active member of the DELPHI collaboration and its QCD
and WW working groups. The loss of Frans touched us all deeply.
}
\end{center}

\end{shortabs}
\vfill
\begin{center}
\DpSubmit \ \\ 
\end{center}
\vfill
\clearpage
\headsep 10.0pt
\addtolength{\textheight}{10mm}
\addtolength{\footskip}{-5mm}
\begingroup
%
\newcommand{\DpName}[2]{\hbox{#1$^{\ref{#2}}$},\hfill}
\newcommand{\DpNameTwo}[3]{\hbox{#1$^{\ref{#2},\ref{#3}}$},\hfill}
\newcommand{\DpNameThree}[4]{\hbox{#1$^{\ref{#2},\ref{#3},\ref{#4}}$},\hfill}
\newskip\Bigfill \Bigfill = 0pt plus 1000fill
\newcommand{\DpNameLast}[2]{\hbox{#1$^{\ref{#2}}$}\hspace{\Bigfill}}
%
\footnotesize
\noindent
\DpName{J.Abdallah}{LPNHE}
\DpName{P.Abreu}{LIP}
\DpName{W.Adam}{VIENNA}
\DpName{P.Adzic}{DEMOKRITOS}
\DpName{Z.Albrecht}{KARLSRUHE}
\DpName{T.Alderweireld}{AIM}
\DpName{R.Alemany-Fernandez}{CERN}
\DpName{T.Allmendinger}{KARLSRUHE}
\DpName{P.P.Allport}{LIVERPOOL}
\DpName{U.Amaldi}{MILANO2}
% ...
\normalsize
\endgroup
\titlefoot{Department of Physics and Astronomy, Iowa State
     University, Ames IA 50011-3160, USA
    \label{AMES}}
\titlefoot{Physics Department, Universiteit Antwerpen,
     Universiteitsplein 1, B-2610 Antwerpen, Belgium \\
     \indent~~and IIHE, ULB-VUB,
     Pleinlaan 2, B-1050 Brussels, Belgium \\
     \indent~~and Facult\'e des Sciences,
     Univ. de l'Etat Mons, Av. Maistriau 19, B-7000 Mons, Belgium
    \label{AIM}}
\titlefoot{Physics Laboratory, University of Athens, Solonos Str.
     104, GR-10680 Athens, Greece
    \label{ATHENS}}
\titlefoot{Department of Physics, University of Bergen,
     All\'egaten 55, NO-5007 Bergen, Norway
    \label{BERGEN}}
\titlefoot{Dipartimento di Fisica, Universit\`a di Bologna and INFN,
     Via Irnerio 46, IT-40126 Bologna, Italy
    \label{BOLOGNA}}
\titlefoot{Centro Brasileiro de Pesquisas F\'{\i}sicas, rua Xavier Sigaud 150,
     BR-22290 Rio de Janeiro, Brazil \\
     \indent~~and Depto. de F\'{\i}sica, Pont. Univ. Cat\'olica,
     C.P. 38071 BR-22453 Rio de Janeiro, Brazil \\
     \indent~~and Inst. de F\'{\i}sica, Univ. Estadual do Rio de Janeiro,
     rua S\~{a}o Francisco Xavier 524, Rio de Janeiro, Brazil
    \label{BRASIL}}
\titlefoot{Coll\`ege de France, Lab. de Physique Corpusculaire, IN2P3-CNRS,
     FR-75231 Paris Cedex 05, France
    \label{CDF}}
\titlefoot{CERN, CH-1211 Geneva 23, Switzerland
    \label{CERN}}
\titlefoot{Institut de Recherches Subatomiques, IN2P3 - CNRS/ULP - BP20,
     FR-67037 Strasbourg Cedex, France
    \label{CRN}}
\titlefoot{Now at DESY-Zeuthen, Platanenallee 6, D-15735 Zeuthen, Germany
    \label{DESY}}
\titlefoot{Institute of Nuclear Physics, N.C.S.R. Demokritos,
     P.O. Box 60228, GR-15310 Athens, Greece
    \label{DEMOKRITOS}}
\titlefoot{FZU, Inst. of Phys. of the C.A.S. High Energy Physics Division,
     Na Slovance 2, CZ-180 40, Praha 8, Czech Republic
    \label{FZU}}
\titlefoot{Dipartimento di Fisica, Universit\`a di Genova and INFN,
     Via Dodecaneso 33, IT-16146 Genova, Italy
    \label{GENOVA}}
\titlefoot{Institut des Sciences Nucl\'eaires, IN2P3-CNRS, Universit\'e
     de Grenoble 1, FR-38026 Grenoble Cedex, France
    \label{GRENOBLE}}
\titlefoot{Helsinki Institute of Physics and  Department of Physical Sciences, P.O. Box 64,
     FIN-00014 University of Helsinki,\\
     \indent~~Finland 
    \label{HELSINKI}}
\titlefoot{Joint Institute for Nuclear Research, Dubna, Head Post
     Office, P.O. Box 79, RU-101 000 Moscow, Russian Federation
    \label{JINR}}
\titlefoot{Institut f\"ur Experimentelle Kernphysik,
     Universit\"at Karlsruhe, Postfach 6980, DE-76128 Karlsruhe,
     Germany
    \label{KARLSRUHE}}
\titlefoot{Institute of Nuclear Physics,Ul. Kawiory 26a,
     PL-30055 Krakow, Poland
    \label{KRAKOW1}}
\titlefoot{Faculty of Physics and Nuclear Techniques, University of Mining
     and Metallurgy, PL-30055 Krakow, Poland
    \label{KRAKOW2}}
\titlefoot{Universit\'e de Paris-Sud, Lab. de l'Acc\'el\'erateur
     Lin\'eaire, IN2P3-CNRS, B\^{a}t. 200, FR-91405 Orsay Cedex, France
    \label{LAL}}
\titlefoot{School of Physics and Chemistry, University of Lancaster,
     Lancaster LA1 4YB, UK
    \label{LANCASTER}}
\titlefoot{LIP, IST, FCUL - Av. Elias Garcia, 14-$1^{o}$,
     PT-1000 Lisboa Codex, Portugal
    \label{LIP}}
\titlefoot{Department of Physics, University of Liverpool, P.O.
     Box 147, Liverpool L69 3BX, UK
    \label{LIVERPOOL}}
\titlefoot{Dept. of Physics and Astronomy, Kelvin Building,
     University of Glasgow, Glasgow G12 8QQ
    \label{GLASGOW}}
\titlefoot{LPNHE, IN2P3-CNRS, Univ.~Paris VI et VII, Tour 33 (RdC),
     4 place Jussieu, FR-75252 Paris Cedex 05, France
    \label{LPNHE}}
\titlefoot{Department of Physics, University of Lund,
     S\"olvegatan 14, SE-223 63 Lund, Sweden
    \label{LUND}}
\titlefoot{Universit\'e Claude Bernard de Lyon, IPNL, IN2P3-CNRS,
     FR-69622 Villeurbanne Cedex, France
    \label{LYON}}
\titlefoot{Dipartimento di Fisica, Universit\`a di Milano and INFN-MILANO,
     Via Celoria 16, IT-20133 Milan, Italy
    \label{MILANO}}
\titlefoot{Dipartimento di Fisica, Univ. di Milano-Bicocca and
     INFN-MILANO, Piazza della Scienza 2, IT-20126 Milan, Italy
    \label{MILANO2}}
\titlefoot{IPNP of MFF, Charles Univ., Areal MFF,
     V Holesovickach 2, CZ-180 00, Praha 8, Czech Republic
    \label{NC}}
\titlefoot{NIKHEF, Postbus 41882, NL-1009 DB
     Amsterdam, The Netherlands
    \label{NIKHEF}}
\titlefoot{Northwestern University, Evanston, IL-60208, USA
    \label{NUE}}
\titlefoot{National Technical University, Physics Department,
     Zografou Campus, GR-15773 Athens, Greece
    \label{NTU-ATHENS}}
\titlefoot{Physics Department, University of Oslo, Blindern,
     NO-0316 Oslo, Norway
    \label{OSLO}}
\titlefoot{Dpto. Fisica, Univ. Oviedo, Avda. Calvo Sotelo
     s/n, ES-33007 Oviedo, Spain
    \label{OVIEDO}}
\titlefoot{Department of Physics, University of Oxford,
     Keble Road, Oxford OX1 3RH, UK
    \label{OXFORD}}
\titlefoot{Dipartimento di Fisica, Universit\`a di Padova and
     INFN, Via Marzolo 8, IT-35131 Padua, Italy
    \label{PADOVA}}
\titlefoot{Rutherford Appleton Laboratory, Chilton, Didcot
     OX11 OQX, UK
    \label{RAL}}
\titlefoot{Dipartimento di Fisica, Universit\`a di Roma II and
     INFN, Tor Vergata, IT-00173 Rome, Italy
    \label{ROMA2}}
\titlefoot{Dipartimento di Fisica, Universit\`a di Roma III and
     INFN, Via della Vasca Navale 84, IT-00146 Rome, Italy
    \label{ROMA3}}
\titlefoot{DAPNIA/Service de Physique des Particules,
     CEA-Saclay, FR-91191 Gif-sur-Yvette Cedex, France
    \label{SACLAY}}
\titlefoot{Instituto de Fisica de Cantabria (CSIC-UC), Avda.
     los Castros s/n, ES-39006 Santander, Spain
    \label{SANTANDER}}
\titlefoot{Inst. for High Energy Physics, Serpukov
     P.O. Box 35, Protvino, (Moscow Region), Russian Federation
    \label{SERPUKHOV}}
\titlefoot{J. Stefan Institute, Jamova 39, SI-1000 Ljubljana, Slovenia
     and Laboratory for Astroparticle Physics,\\
     \indent~~Nova Gorica Polytechnic, Kostanjeviska 16a, SI-5000 Nova Gorica, Slovenia, \\
     \indent~~and Department of Physics, University of Ljubljana,
     SI-1000 Ljubljana, Slovenia
    \label{SLOVENIJA}}
\titlefoot{Fysikum, Stockholm University,
     Box 6730, SE-113 85 Stockholm, Sweden
    \label{STOCKHOLM}}
\titlefoot{Institute of High Energy Physics of Tiblisi State University, Georgia
    \label{TIB}}
\titlefoot{Dipartimento di Fisica Sperimentale, Universit\`a di
     Torino and INFN, Via P. Giuria 1, IT-10125 Turin, Italy
    \label{TORINO}}
\titlefoot{INFN,Sezione di Torino, and Dipartimento di Fisica Teorica,
     Universit\`a di Torino, Via P. Giuria 1,\\
     \indent~~IT-10125 Turin, Italy
    \label{TORINOTH}}
\titlefoot{Dipartimento di Fisica, Universit\`a di Trieste and
     INFN, Via A. Valerio 2, IT-34127 Trieste, Italy \\
     \indent~~and Istituto di Fisica, Universit\`a di Udine,
     IT-33100 Udine, Italy
    \label{TU}}
\titlefoot{Univ. Federal do Rio de Janeiro, C.P. 68528
     Cidade Univ., Ilha do Fund\~ao
     BR-21945-970 Rio de Janeiro, Brazil
    \label{UFRJ}}
\titlefoot{Department of Radiation Sciences, University of
     Uppsala, P.O. Box 535, SE-751 21 Uppsala, Sweden
    \label{UPPSALA}}
\titlefoot{IFIC, Valencia-CSIC, and D.F.A.M.N., U. de Valencia,
     Avda. Dr. Moliner 50, ES-46100 Burjassot (Valencia), Spain
    \label{VALENCIA}}
\titlefoot{Institut f\"ur Hochenergiephysik, \"Osterr. Akad.
     d. Wissensch., Nikolsdorfergasse 18, AT-1050 Vienna, Austria
    \label{VIENNA}}
\titlefoot{Inst. Nuclear Studies and University of Warsaw, Ul.
     Hoza 69, PL-00681 Warsaw, Poland
    \label{WARSZAWA}}
\titlefoot{Fachbereich Physik, University of Wuppertal, Postfach
     100 127, DE-42097 Wuppertal, Germany
    \label{WUPPERTAL}}
\addtolength{\textheight}{-10mm}
\addtolength{\footskip}{5mm}
\clearpage
\headsep 30.0pt
\end{titlepage}
%
\pagenumbering{arabic} 
\setcounter{footnote}{0} %
\large
\section{Introduction}
\label{sec:sec1}
Correlations between final-state particles in high energy collisions have
been extensively studied during the last decades. They can be due to phase space, energy-momentum
conservation, resonance production, hadronisation mechanisms, or be
dynamical in nature.

In the particular case of identical bosons the correlations 
are enhanced by the Bose-Einstein effect~\cite{gold1,gold2}.
These Bose-Einstein correlations (BEC) are a consequence of quantum
statistics. The net result is that multiplets of
identical bosons are produced with smaller energy-momentum differences than
non-identical ones.

Several aspects of BEC have been measured in hadronic Z decays and are
well understood~\cite{LEPBECZ0}. It is  natural to expect the
same behaviour in the hadronic decay of a single W. It is, however, not clear
how BEC manifest themselves in a system of two hadronically decaying W's, in
particular between bosons coming from different W's (inter-W BEC).

The separation between two W's before their decay is of the order of 0.1 fm, 
compared to a typical hadronisation scale of
several fm. Therefore, due to the large overlap between the two hadron
sources, inter-W BEC cannot be a priori excluded.
However, it is unclear whether these are of the same 
type as BEC measured inside a single decaying W, where they are, in contrast to the
traditional Hanbury-Brown and Twiss~\cite{hbt1} picture, not related to the total hadronisation volume.

Together with colour reconnection~\cite{skmodel,arcr}, the poor understanding of the
inter-W BEC effect introduces a large systematic uncertainty in the measurement
of the W mass in the channel  $\rm{e^+e^- \to W^+W^- \to
q_1\overline{q_2}q_3\overline{q_4}}$~\cite{Lonnblad:1995mr,appref}.
The current statistical uncertainty of the combined LEP measurement in this
channel amounts to 35 MeV~\cite{lepewnote}, to be compared with the total systematic
uncertainty in this channel of 107 MeV, which is, however, expected to decrease with
improved measurements of colour reconnection. The effect of possible
inter-W BEC amounts to 35 MeV~\cite{lepewnote}. It is thus clear that a better
understanding of the phenomenon would help in reducing this
uncertainty.

Measuring inter-W BEC is challenging in practice because of a low
sensitivity to the effect. This is mainly due to the small fraction of
relevant particle pairs coming from different W's. Moreover, its isolation 
from BEC inside a single W requires careful attention and needs to be 
as model-independent as possible.

The scope of this paper is the model-independent analysis of the
correlations
of like-sign hadron pairs in $\rm{e^+e^- \to W^+W^-}$, where both W's
decay into hadrons, with the aim of determining the presence and size of
inter-W BEC.

The outline of the paper is as follows: In section~\ref{sec:sec2} the
mathematical
formalism applied throughout this analysis is specified and a brief
overview
of the analysis is given. In section~\ref{sec:sec3} experimental details,
such as the detector setup and WW event selection, are presented. Section~\ref{sec:sec4}
focuses on the mixing procedure employed in order to construct an inter-W
BEC-free reference sample from events where one W decays leptonically.
Section~\ref{sec:sec5} clarifies the details of the Monte Carlo models for
comparison to the data. 
In section~\ref{sec:sec6} a detailed
overview
of the numerical analysis of the measured correlation functions is given
including a construction of weights applied in order to increase the
sensitivity of the analysis, the subtraction of background and the
determination of statistical errors and correlations of the bins.
Moreover the parametrisation of the correlation function is discussed.
In section~\ref{sec:sec7} results are presented and in section~\ref{sec:sec8} the
systematic
uncertainties are discussed. Finally, sections~\ref{sec:sec9}
and~\ref{sec:sec10} discuss the results and conclusions are given.
\section{Analysis method}
\label{sec:sec2}
The mathematical method used to extract a possible 
inter-W BEC signal is largely based on
\cite{Chekanov:1998hi} and \cite{DeWolf:2001wz}. In
the case of two stochastically independent hadronically decaying W's, 
 the single and two-particle inclusive densities obey the following relations:

\begin{align}
\rho^{\rm WW}(1) &= \rho^{\rm W^+}(1)+\rho^{\rm W^-}(1), \\
\rho^{\rm WW}(1,2) &= \rho^{\rm W^+}(1,2)+\rho^{\rm W^-}(1,2)+
\rho^{\rm W^+}(1)\rho^{\rm W^-}(2)
+\rho^{\rm W^+}(2)\rho^{\rm W^-}(1),
\label{densities}
\end{align}  

\noindent
where $\rho^{\rm W}(1)$ denotes the inclusive single particle density of one W and $\rho^{\rm W}(1,2)$
the inclusive two-particle density of one W. The densities $\rho^{\rm WW}(1)$ and
$\rho^{\rm WW}(1,2)$ then correspond to the single and two-particle inclusive
densities of a fully-hadronic WW event.
Assuming that the densities for the $\rm{\rm W^+}$ and the $\rm{\rm W^-}$ are the same,
which is correct if one does not look at the absolute sign of the particles'
charges, equation~(\ref{densities}) can be re-written as

\begin{equation}
\rho^{\rm WW}(1,2) = 2\rho^{\rm W}(1,2)+2\rho^{\rm W}(1)\rho^{\rm W}(2).
\label{simpler}
\end{equation}

The terms $\rho^{\rm WW}(1,2)$ and $\rho^{\rm W}(1,2)$ can be measured in
 fully-hadronic and semi-leptonic WW decays respectively. 

A pair or two-particle density $\rho^{\rm WW}(1,2)$ is trivial to construct.
The correlation measurement is made difficult by the fact that
only the correlations between particles coming from different W's
are of interest and there is no way of determining where the particles 
originated from. Finally, in order to obtain a correlation function, it is necessary
to construct a reference sample of events without BEC between particles
 coming from different W bosons. This sample corresponds, in our case, to the product of the
single particle densities $\rho^{\rm W}(1)\rho^{\rm W}(2)$.

Events where only one of the W's decays hadronically can be used 
to address these challenges. 
Taking two of these independent 
hadronically decaying W's and mixing them to form one event allows an 
emulation of a fully-hadronic WW event, having BEC inside each of the W's.
By construction these events 
will have no correlations in pairs from different W's and the
measurement becomes a direct comparison between two event samples,
without any model dependence. The event mixing should follow closely the electroweak production of WW
events. Possible biases of the mixing procedure can be estimated by
applying the same procedure to large samples of simulated events.
Hence, the term $\rho^{\rm W}(1)\rho^{\rm W}(2)$ in equation~(\ref{simpler}) 
replaced by a
two-particle density $\rho^{\rm WW}_{\rm mix}$, obtained by combining particles
from two hadronic W decays taken from different semi-leptonic events. 
The details of this ``mixing'' procedure are explained in 
section~\ref{sec:sec4}.
Expressed in the variable $Q=\sqrt{-(p_1-p_2)^2}$, where $p_{1,2}$ stands for
the four-momentum of particles $1$ and $2$, equation~(\ref{simpler}) can be
re-written as

\begin{equation}
\rho^{\rm WW}(Q) = 2\rho^{\rm W}(Q)+2\rho^{\rm WW}_{\rm mix}(Q).
\label{msimpler}
\end{equation}

Keeping in mind that equation~(\ref{densities}) was formulated for
independent W decays, test observables can be constructed to search for deviations
from this assumption. Such deviations will indicate that particles from different W decays do
correlate. The observables considered are:

\begin{align}
\Delta \rho (Q) &= \rho^{\rm WW}(Q) - 2\rho^{\rm W}(Q) - 2\rho^{\rm WW}_{\rm mix}(Q), \label{drho}\\
D(Q) &= \frac{\rho^{\rm WW}(Q)}{2\rho^{\rm W}(Q) + 2\rho^{\rm WW}_{\rm
    mix}(Q)}. \label{d}\\
\intertext{Given the definition of the genuine inter-W correlation function
  $\delta_I(Q)$~\cite{DeWolf:2001wz}, it can be shown that}
\delta_I(Q) &= \frac{\Delta \rho (Q)}{2\rho^{\rm WW}_{\rm mix}(Q)}. \label{deli}
\end{align}   
%

If no inter-W correlations exist, the variables $\Delta \rho (Q)$ and $\delta_I(Q)$ will be zero
for all values of $Q$, while $D(Q)$ will be equal to one.
Inter-W BEC will lead to an excess at small values of $Q$.

The selection of particles and pairs is straightforward, with
the strongest requirement that they should originate from the primary interaction.
Moreover, the selected WW candidates have a significant background which
must be subtracted using a model dependent procedure.

BEC in Z decays have been extensively measured and constitute a 
natural basis to compare with inter-W BEC. The correlation functions 
measured in Z events use simulated events without BEC as reference 
samples. They are therefore close to being genuine correlation 
functions but with large model-dependent systematic errors and
some dilution due to particles which are either not pions or which 
do no originate from the primary interaction. When the inter-W 
correlations are measured it is natural to compare to the Z and 
single-W data using the same fitting functions. Since the inter-W
measurement uses data as reference, the model dependence
is no longer present.

The mixing procedure, which allows events to be mixed more than once,
leads to a rather involved description of the statistical properties 
of the correlation function.
%
However, the same mixing can be used to investigate the sensitivity 
to the inter-W BEC effect. 
The applied mixing reuses semi-leptonic events up to 20 times,
which affects the precision depending on whether pairs 
are constructed by mixing or come from inside single W's.
Finally, a pair-weighting technique was devised which improved the
sensitivity and is described in section~\ref{sec:sec6.1}.
For this purpose the mixed reference sample was 
used to determine statistically whether particles come 
from the same or different W's. 
\section{Experimental details}
\label{sec:sec3}
\subsection{The DELPHI detector}
The DELPHI detector configuration for the LEP2 running evolved 
compared to the one at LEP1~\cite{Aarnio:1991vx,Abreu:1996uz}.
The main changes relevant to the analysis described in this paper
were the extension of both the vertex and
the inner detectors. This ensured a very good track quality
also in the forward region down to small polar angles.
During the operation of the detector in the latter part of the year 2000 one sector of the TPC
malfunctioned and the data from this period are excluded from the 
results.

In order to verify that a track originates from the primary 
interaction it was required that the TPC participated in the measurement of
the track.  This effectively required 
the track to be within the polar angle region 
$20$\degr $< \theta < 160$\degr. 
The reconstructed charged particles were required
to fulfill the following criteria on the momentum, $p$, the momentum error,
$\Delta p / p$, and the impact parameters with respect to the event vertex in
the plane transverse to the beam, $\epsilon_{\perp}$, or parallel to the
beam, $\epsilon_{\parallel}$:

\begin{itemize}
\item 0.2 GeV $< p < p_{\rm beam}$;
\item $\Delta p / p < 1$;
\item $\epsilon_{\perp} < 0.4$ cm;
\item $\epsilon_{\parallel} < 1.0$ cm/sin$\theta$.
\end{itemize}

%
%
The two track reconstruction efficiency in DELPHI drops 
for opening angles below 2.5$^{\circ}$. Since the mixing procedure does not
necessarily reproduce this drop in efficiency all particle pairs having an
opening angle below 2.5$^{\circ}$ were omitted in all two-particle density distributions.
These requirements lead to a typical efficiency of about
85\%
and reduce the total fraction of secondary tracks to about 5\%.
Secondary tracks are typically tracks from
  secondary decays ($K^0$, $\Lambda^0$, etc.) or from secondary interactions
  in the beam pipe and with detector material.
Particles not coming from the primary interaction or not being pions will
dilute the observed correlation.
The combined effect was estimated to reduce the measured BEC to about 70\% of
the nominal one. This dilution was not corrected for due to model dependence
and affects all pair densities in nearly the same way. When results from
different experiments are combined it will be necessary to apply such
corrections in order to get comparable results.

\subsection{Selection of WW events}
\begin{table}[!t]
\begin{center}
\begin{tabular}{|c||c|c|c|c|c|c|} \hline
Year & 1998 & \multicolumn{4}{|c|}{1999}
&2000
\cr
\hline\hline
$\sqrt{s}$ (GeV)&  189 & 192 &196 &200 &202 & 204-209 \cr
$\mathcal{L}$ (pb$^{-1}$) & 158.0 & 25.9 & 76.9 & 84.3 & 41.1 & 163.4 \cr
\hline
\end{tabular}
\caption{The integrated luminosities, $\mathcal{L}$, for the various years of
  LEP2 data-taking, expressed in units of pb$^{-1}$. The corresponding
  centre-of-mass energies are also given.}
\label{lumitab}
\end{center}
\end{table}
The total analysed dataset amounts to an integrated luminosity of 549.6 pb$^{-1}$,
collected with the DELPHI detector during the years 1998--2000. 
A summary of  the integrated luminosity per energy point is
given in table~\ref{lumitab}.

The samples of fully-hadronic and semi-leptonic events required for
the WW BEC
analysis were selected using neural networks, developed in~\cite{crosspaper}
and~\cite{chhiggs}.  

For the {\bf fully-hadronic} event selection, it was demanded that the events fulfill the
following requirements:
a large enough charge multiplicity, 
a large effective centre-of-mass energy, large visible energy and
four or more jets.

\begin{figure}[!tb]
\hspace{-13. mm}
  \begin{tabular}{cc}
      \epsfxsize=8.5 cm 
\epsfbox{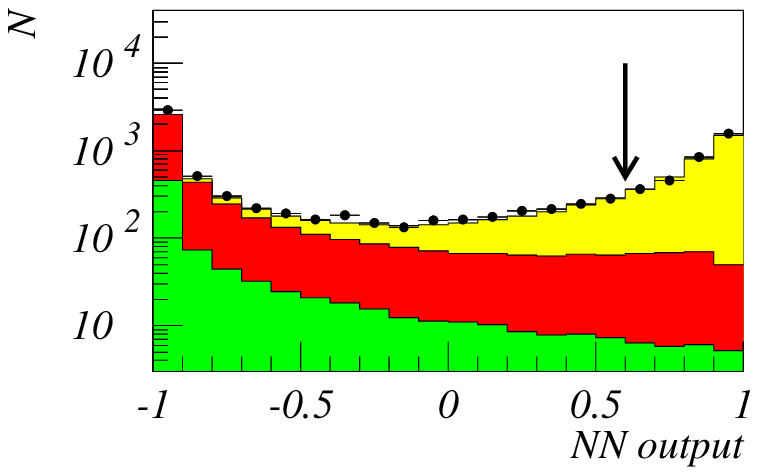}
&
      \epsfxsize=7.5 cm 
\epsfbox{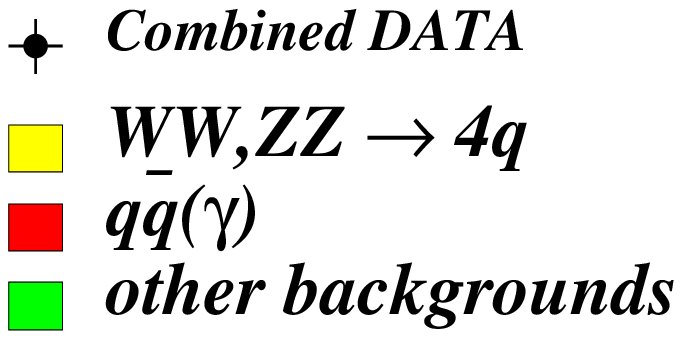}
  \end{tabular}
  \caption{The neural network output variable for the fully-hadronic
  event selection. The light shaded histogram are the signal,
  while the dark shaded histograms correspond to the background
  processes. The optimised selection cut is indicated by the arrow.}
  \label{nnoutfh} 
\end{figure}

The final selection was performed using a neural network trained on thirteen
event variables.
The dominant background contribution
comes from
the $\rm{q\overline{q}(\gamma)}$ events. All other backgrounds are
negligible. 
Hadronically decaying ZZ events, which constitute 5\% of the selected
sample, were treated as signal as they, except for events where at least
one Z decays into b-quarks, will have similar space-time kinematics.
A comparison between data and simulated events of the neural network output 
for the fully-hadronic selection is shown in figure~\ref{nnoutfh}.

By requiring a neural network output larger than a given value, a desired
purity or efficiency can be reached.
The whole analysis was repeated for several cuts on the neural network output,
selecting samples with an increasing purity, ranging from 83\% to 97\%. 
This allowed the choice of an optimal working point, minimising the 
sum of the statistical and background uncertainty, corresponding to a
selection efficiency and purity of 63\% and 92\% respectively, with 3252
events selected in total. 

\begin{figure}[!bth]
\hspace{-13. mm}
  \begin{tabular}{cc}
      \epsfxsize=8.5 cm 
\epsfbox{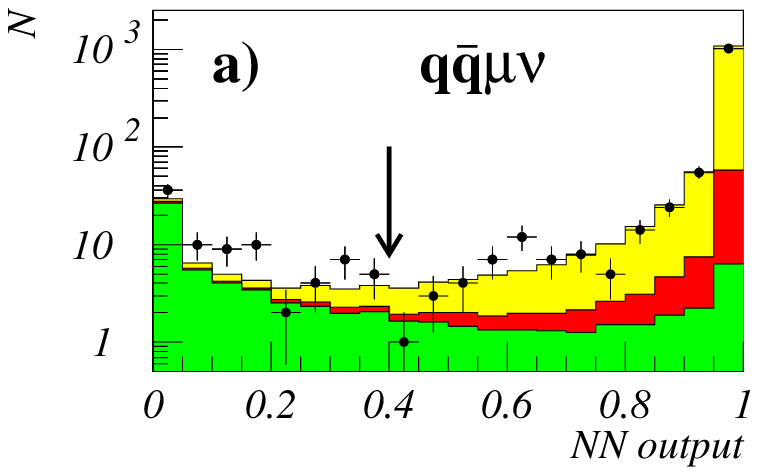}
&
      \epsfxsize=8.5 cm 
\epsfbox{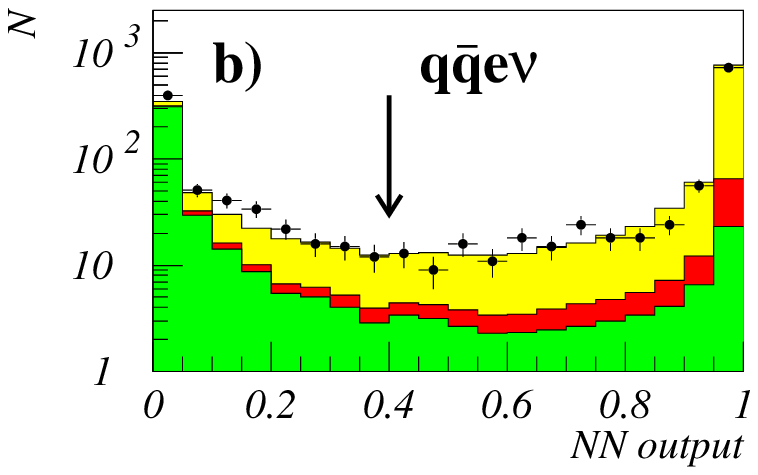}
\\
      \epsfxsize=8.5 cm 
\epsfbox{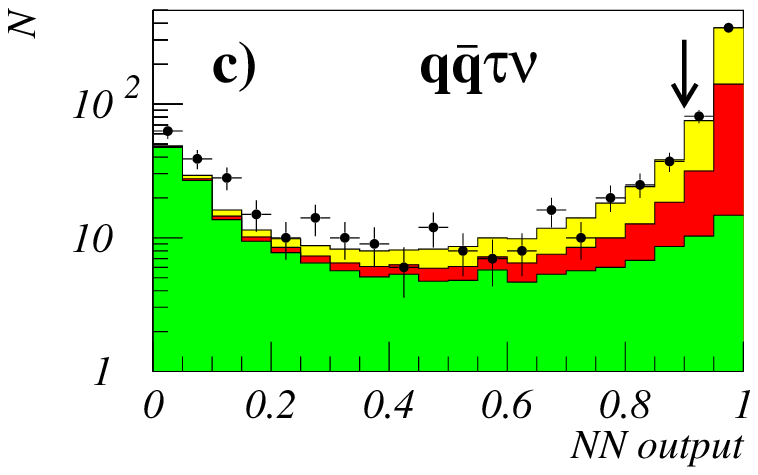}
      & 
      \epsfxsize=8.5 cm 
\epsfbox{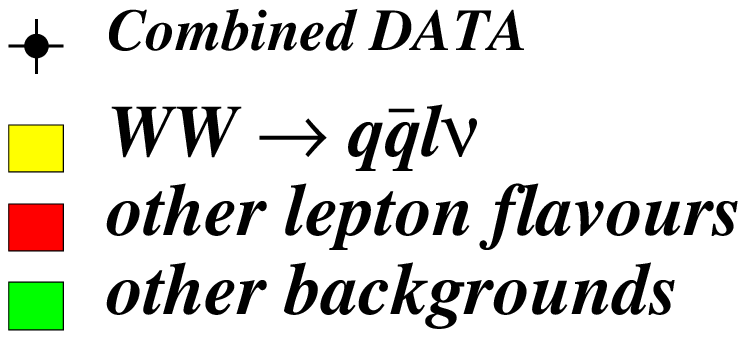}
  \end{tabular}
  \caption{The neural network outputs for the three semi-leptonic
  event selections. The muon channel (a) and electron channel (b)
  have small background
  contaminations, due to the clear identification of the isolated
  lepton. (c) The taus are more difficult to identify, resulting in a
  higher background rate. All signal events are shown by the light
  shaded histograms, the background events correspond to the dark
  shaded histograms. The selection cuts are indicated by the arrows.}
  \label{nnoutsl} 
\end{figure}
The {\bf semi-leptonic} events were selected by requiring two hadronic jets, a well-isolated
identified muon or electron or (for tau candidates) a well-isolated particle
associated with missing momentum possibly from the neutrino.
The missing momentum direction was required to point away from the beam pipe.
Dedicated neural network trainings were used for all  lepton flavours. 
Combining all three lepton flavours, an overall efficiency  and purity of respectively 58\% and 96\%
was reached, corresponding to 2567 selected events. The three neural network
outputs, corresponding to the three lepton flavours are shown in
figure~\ref{nnoutsl} for data and simulated events.

The WPHACT~\cite{wphact} generator with
the JETSET~\cite{Sjostrand:1995iq} hadronisation model was used for the
simulation of all signal and
four-fermion background events. The $\rm{q\overline{q}(\gamma)}$ background was
simulated using the KK2F~\cite{kk2f} generator and also hadronized with
JETSET.   
\section{Mixing procedure}
\label{sec:sec4}
\begin{figure}[!t]
\hspace{-13. mm}
  \begin{tabular}{cc}
      \epsfxsize=8.5 cm 
\epsfbox{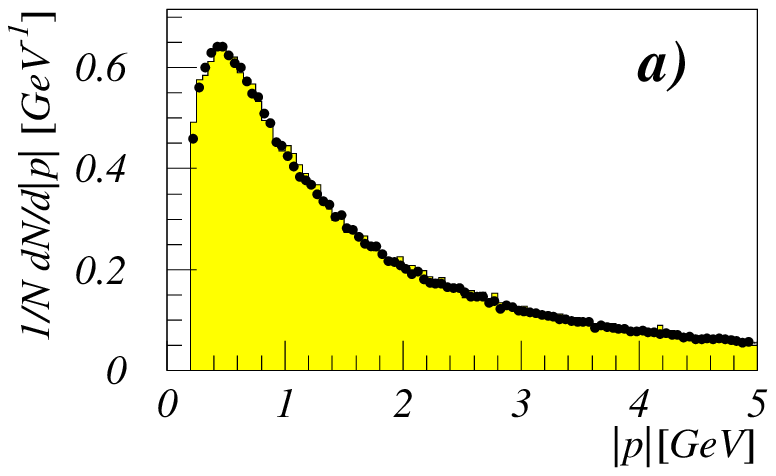}
&
      \epsfxsize=8.5 cm 
\epsfbox{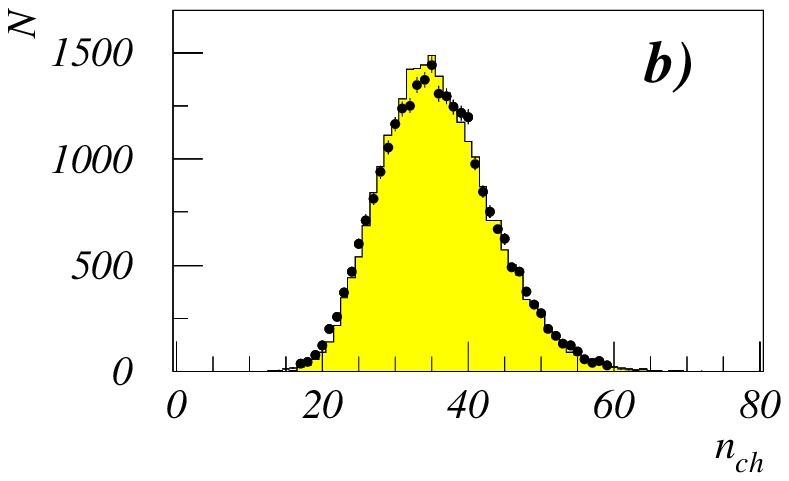}
\\
      \epsfxsize=8.5 cm 
\epsfbox{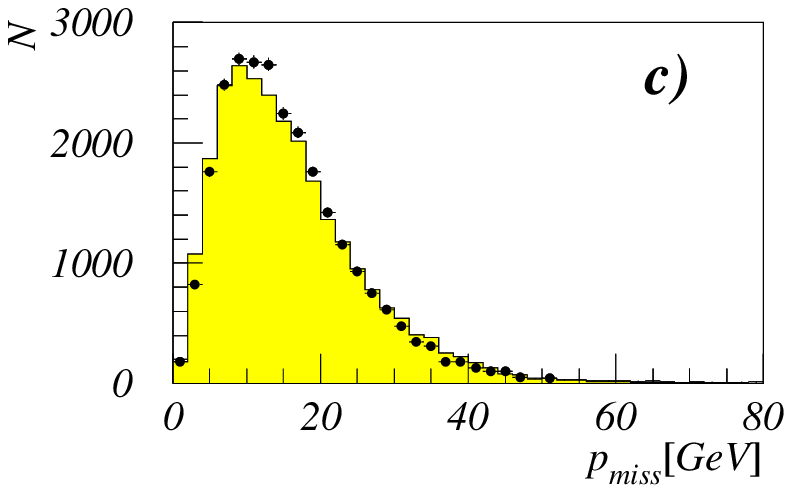}
      & 
      \epsfxsize=8.5 cm 
\epsfbox{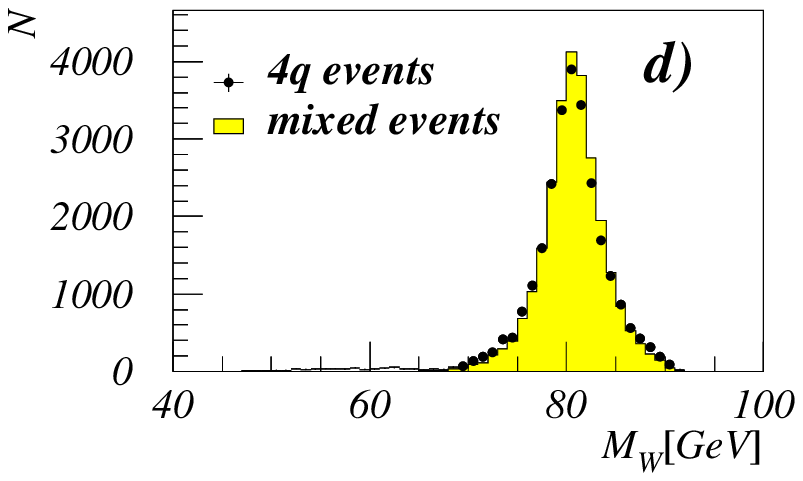}
  \end{tabular}
  \caption{Comparison between simulated fully-hadronic events and simulated mixed events for
  (a) the charged particle momenta, (b) the charge multiplicity, (c) the
  missing momentum and (d) the reconstructed W mass. In all plots the WPHACT
  generator using the BEI model was used.}
  \label{mixqual} 
\end{figure}
%
The mixed two-particle density, $\rho^{\rm WW}_{\rm mix}$, was constructed by combining the hadronically-
decaying W's from pairs of different semi-leptonic WW events,
 $\rm{q\overline{q}l\overline{\nu_l}}$, from which the lepton was removed and
irrespective of the charge of the W's.


The momentum of each hadronic W can be constructed as the
visible W momentum or the W momentum after a constrained fit
imposing energy- and momentum-conservation and constraining the
two W masses in the event to be equal to 80.35 GeV.
After mixing the W's, the first method gives mixed events which
have smaller missing momenta than the fully-hadronic events while the
second gives larger missing momenta. 
It was therefore decided to use the average of the visible W momentum
with a weight of 0.4 and the fitted W momentum
with a weight of 0.6 to obtain the final W momentum. This procedure
gave the best agreement with respect to the missing momenta, and it was 
cross checked that the mixing quality does not depend significantly 
on the used weights.

In WW events the W's are nearly back-to-back due to momentum 
conservation. In mixed events this was accomplished by requiring that 
one W polar angle lay within 10$^{\circ}$ opposite to the polar angle of the
other W.
Pairings where the two W polar angles were within
10$^{\circ}$  after inverting the $z$-component
of one W were also accepted. The momenta of the W's were then approximatively 
balanced by rotating one W around the beam axis so that the 
W's became back-to-back in the plane transverse to the beam axis.
The above transformations reflect the azimuthal and forward-backward symmetry 
of the DELPHI tracking detectors. 

\begin{figure}[!t]
\begin{center}
\mbox{\epsfig{file=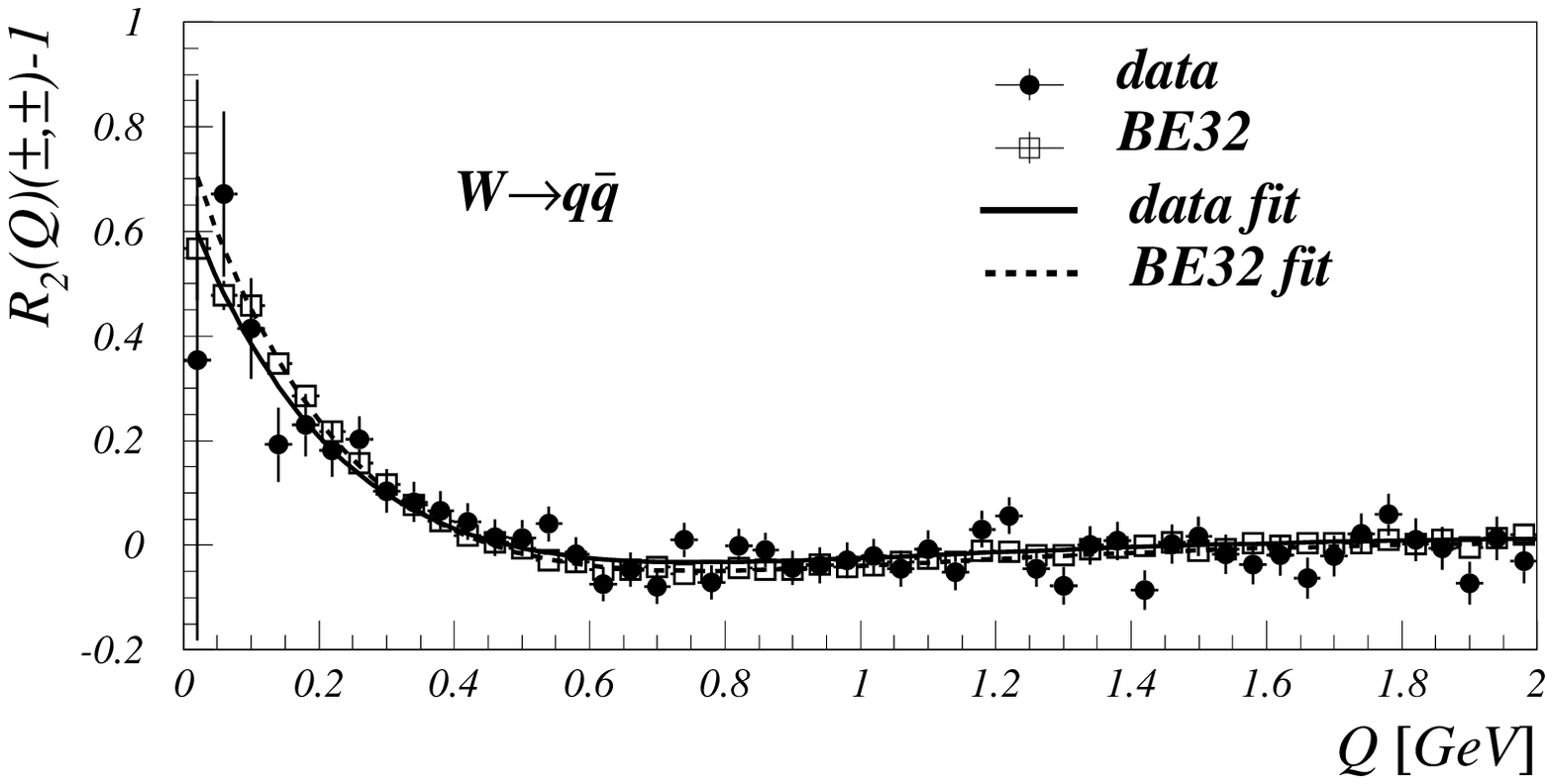,height=6.5cm}}
\caption{The two-particle correlation function for a single W decay, obtained
from semi-leptonic WW events in data and the BE$_{32}$ Monte Carlo model. The
model was tuned using Z data. Fits with
equation~(\ref{laguerre}) are superimposed.}
\label{r2q}
\end{center}
\end{figure}
\noindent

All mixed events were subjected to the same event selection as the
fully-hadronic events. The agreement between fully-hadronic events and mixed events was verified for several
event variables and single-particle distributions. Small differences in the
distributions are taken into account in the estimation of the systematic errors.
Examples are shown in figure~\ref{mixqual}, comparing the particle momenta,
charge multiplicity, total missing momentum and reconstructed W mass between
simulated fully-hadronic events and simulated mixed events.
These events were generated with the BEI model described in section~\ref{sec:sec5}.

\section{Monte Carlo models}
\label{sec:sec5}
All Monte Carlo generated events were hadronized using the JETSET algorithm
unless stated otherwise. BEC were included using the local reweighting
algorithm LUBOEI~\cite{Lonnblad:1998kk,Lonnblad:1995mr}.
It takes as starting point the hadrons produced by the string
fragmentation in JETSET, where no Bose-Einstein effects are present. Then
the momenta of identical bosons are shifted such that the inclusive distribution
of the relative separation $Q$ of identical pairs is enhanced by a factor
$f_2(Q) \geq 1$, parametrised with the phenomenological form
\begin{equation}
f_2(Q)=1+\lambda \exp(-Q^2R^2),
\end{equation}
where $Q$ is the difference in four-momentum of the pair, $\lambda$ and $R$
are free parameters related to the correlation strength and 
the spatial scale of the source of the correlations.

The corresponding change in the momentum of the particles  is not unique. In addition, energy and
momentum cannot be simultaneously conserved. In the model, the momentum
is always conserved and afterwards all three-momenta are rescaled by a
constant factor, close to unity, in order to restore energy conservation.
Even when BEC is only allowed for pairs coming from the same W (BEI), this global
rescaling introduces unreasonable negative shifts in the reconstructed W mass.
The BE$_{32}$ variant of LUBOEI overcomes this problem by including extra momentum
shifts to restore total energy conservation, instead of a rescaling of the
momenta.

The BE$_{32}$ model was tuned to hadronic Z decays, keeping all fragmentation
parameters fixed, giving a satisfactory result for all hadronic Z events and
hadronic Z events with
reduced b-content. The resulting LUBOEI 
parameters for the correlation strength, $\lambda$, and the correlation
length scale, $R$, are 
PARJ(92)=1.35 and PARJ(93)=0.34 GeV$^{-1}$ (= 0.6 fm), respectively.
Monte Carlo sets exceeding ten times the size of the WW data set were simulated at
each centre-of-mass energy.

The tuned model gives a good description of DELPHI's Z data (see
figure~\ref{bgtune}(a)) and the hadronic
decay of single W's. The latter is illustrated in figure~\ref{r2q}, where the 
two-particle correlation function defined as
\begin{equation}
R_2(Q)-1=\frac{\rho^{\rm W}(Q)_{\rm data}}{\rho^{\rm W}(Q)_{\rm MC no BE}}-1,
\end{equation}
is shown for selected semi-leptonic W decays in data and MC simulation.
The dip around $Q=0.5-1.0$ GeV in the BE$_{32}$ curve in figure~\ref{r2q} is 
understood to come from the conservation of the total multiplicity
in the model and is taken into account by the fit in equation
(\ref{laguerre}). 
The signal at low Q values is naturally compensated by
a depletion at higher Q values. 

In this paper the label BEI is used for the LUBOEI model
including only BEC between particles from the same W, BEA is used when
all particles are subjected to BEC and BE0 when all BEC are switched off. 
\begin{figure}[!t]
\begin{center}
\begin{tabular}{cc} 
\hspace{-1.cm}
\mbox{\epsfig{file=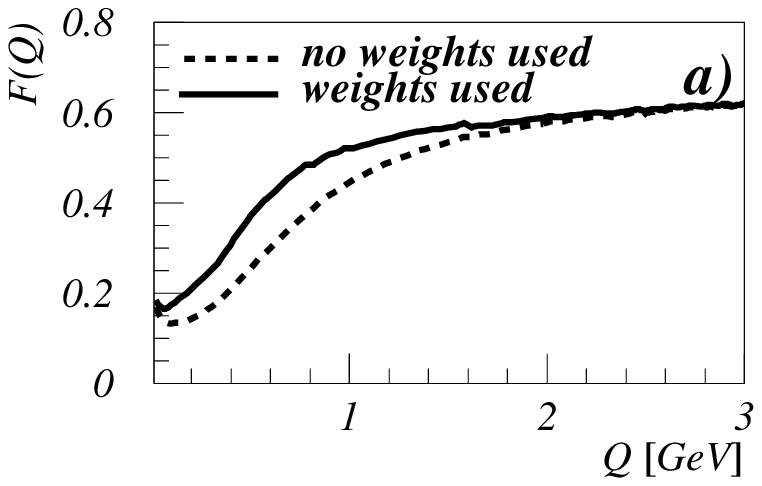,height=5.5cm}} &
\mbox{\epsfig{file=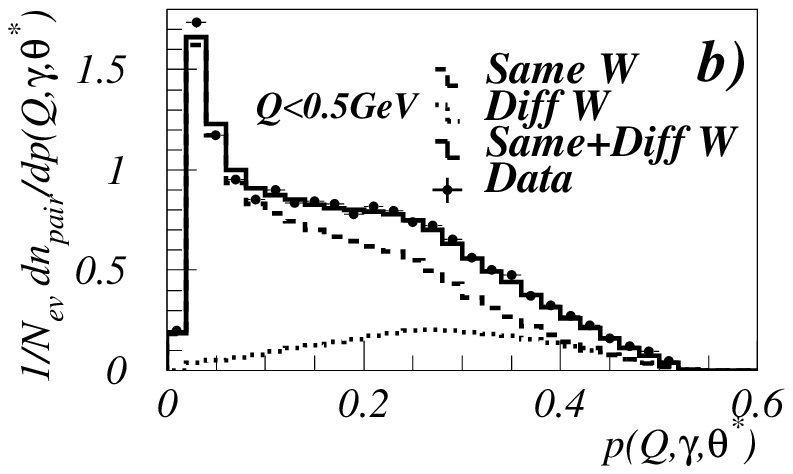,height=5.5cm}}\\
\end{tabular}
\caption{a):The fraction of pairs coming from different $W$'s, $F(Q)$, obtained
  for a BEI MC sample without pair weights (dashed line) and using pair
  weights (full line). b):The estimated pair purity, $p(Q,\gamma,\theta^*)$,
  obtained from mixed semi-leptonic events.}
\label{fofq}
\end{center}
\end{figure}
\begin{figure}[!t]
\begin{center}
\begin{tabular}{cc} 
\hspace{-1.cm}
\mbox{\epsfig{file=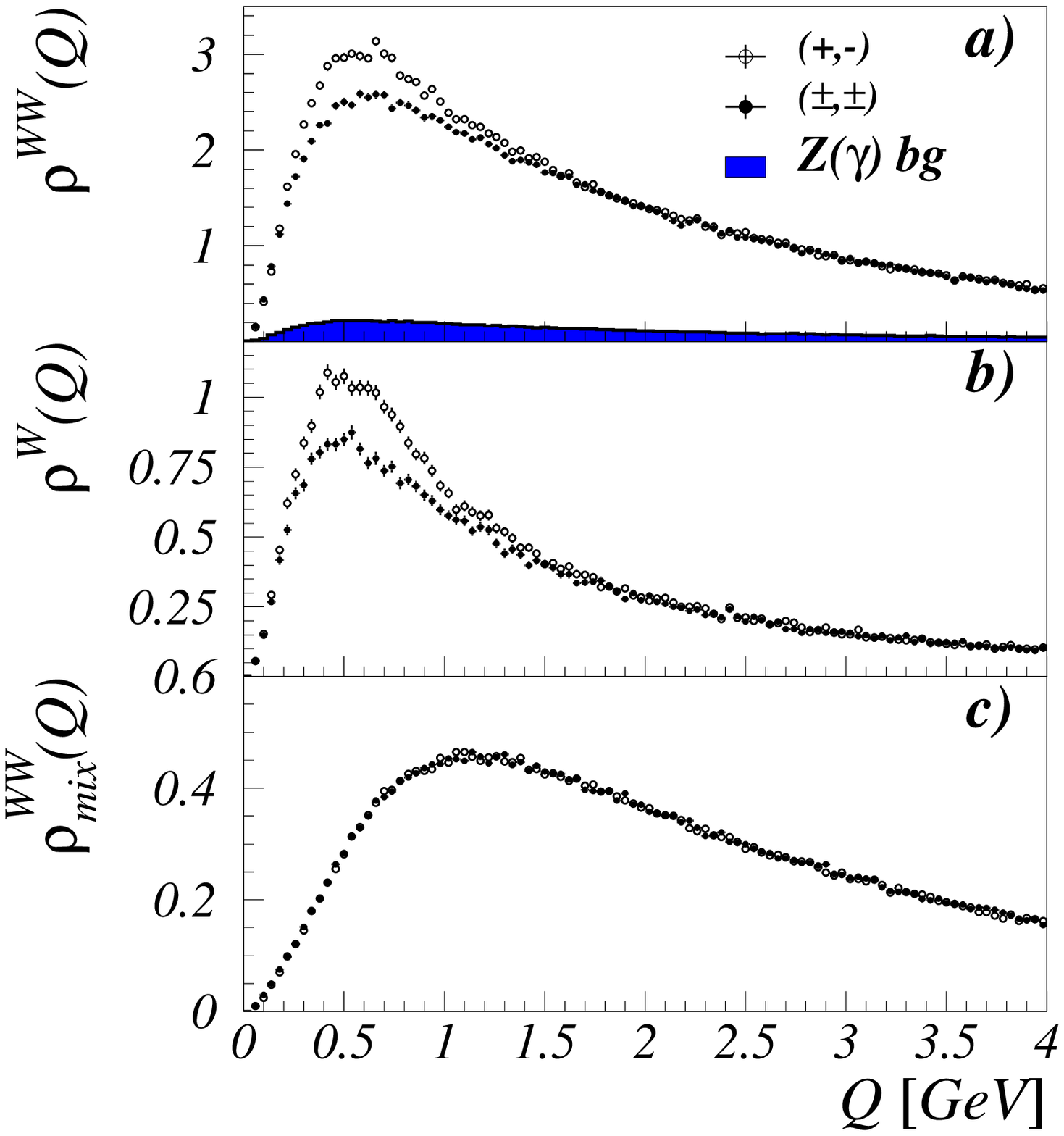,height=9.cm}} &
\mbox{\epsfig{file=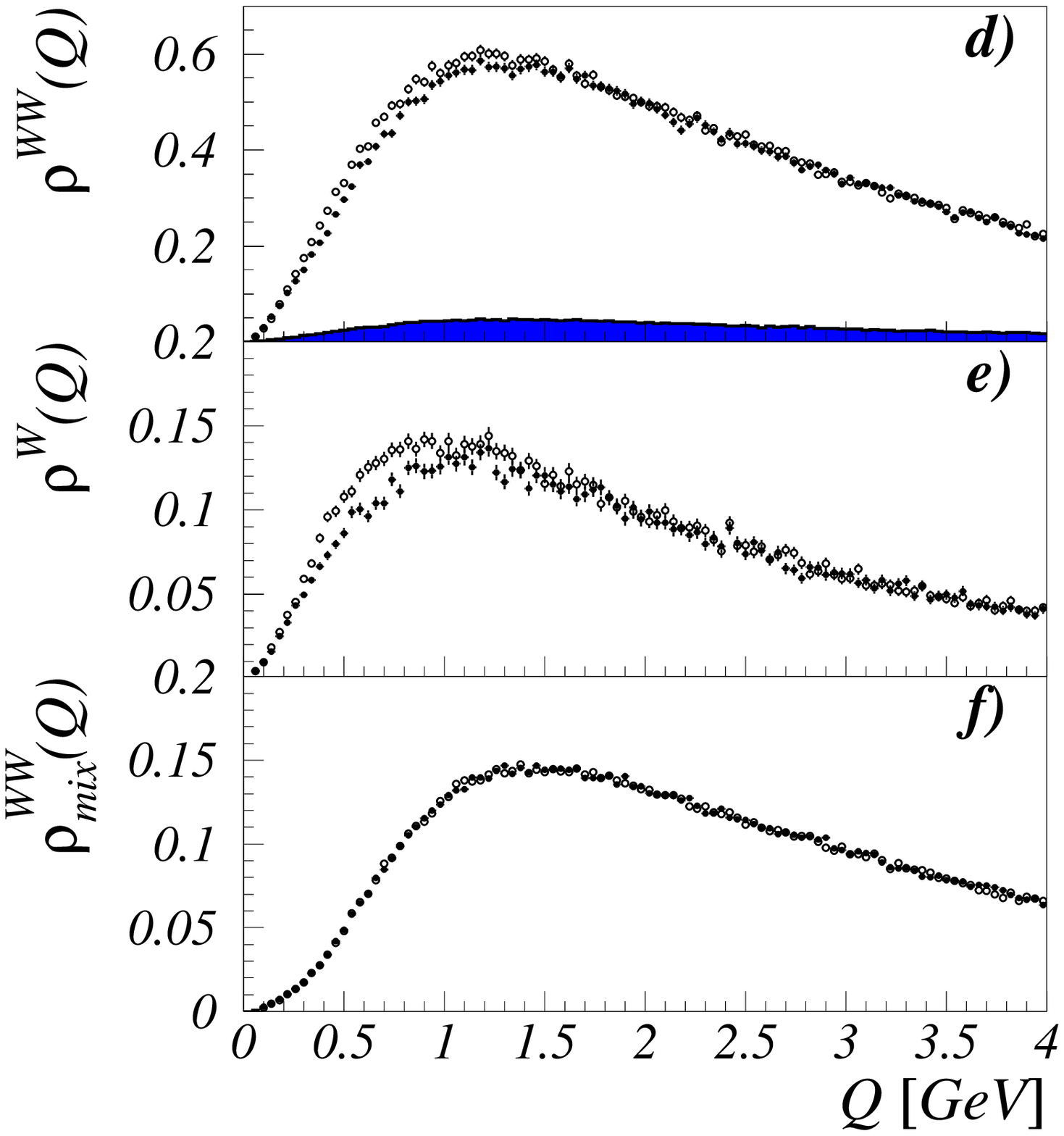,height=9.cm}}\\
\end{tabular}
\caption{The two-particle densities, $\rho^{WW}(Q)$, $\rho^{W}(Q)$ and
  $\rho^{WW}_{mix}(Q)$, for like-sign and unlike-sign particle pairs,
   with no pair weights applied (a-c), and including pair weights (d-f). 
   The background contribution from $\rm{Z(\gamma)\to q\overline{q}(\gamma)}$ events in the $\rho^{WW}(Q)$ distribution
  is also shown.}
\label{qplots}
\end{center}
\end{figure}
\section{Numerical analysis}
\label{sec:sec6}
The numerical analysis of the BEC measurement is complicated by
statistical correlations introduced by the methodology. Each charged particle
occurs in several pairs and each semi-leptonic event is
used to produce several different mixed events. All these
statistical correlations were included in the numerical analysis and
the performance evaluated using resampling techniques.
\subsection{Pair weights}
\label{sec:sec6.1}
The sensitivity of any inter-W BEC measurement is limited by the small
fraction of particle pairs coming from different W's, resulting in a small 
$Q$ value. This is illustrated in  figure~\ref{fofq}(a), where the fraction of
pairs from different W's, denoted as $F(Q)$, is shown.
It drops to around 15\% at very low $Q$ values.

In addition to $Q$, the Lorentz factor, $\gamma$, and the decay angle
of a particle pair, $\theta^*$, are sensitive to whether two particles come
from the same or from different W's. The decay angle, $\theta^*$, is defined
as the angle between the momentum vector of one of the particles in the
two-particle rest frame and the vector sum of the two particle momenta in the
lab frame. As such, each individual pair of tracks can be estimated to have a 
probability $p(Q,\gamma,\theta^*)$ to come from different W's.
$p(Q,\gamma,\theta^*)$ was parametrised for this analysis
using large samples of simulated mixed semi-leptonic W decays.
Figure~\ref{fofq}(b) illustrates the distribution of $p(Q,\gamma,\theta^*)$ for pairs with $Q<0.5$ GeV.
The BEI model is shown for the mixed and same W's and compared to the
data results.

A particle is combined with the other particles in one event
when constructing $\rho^{\rm W}(Q)$. It is combined with 
the other particles in many other events when constructing 
$\rho^{\rm WW}_{\rm mix}(Q)$. Therefore, $\rho^{\rm WW}_{\rm mix}(Q)$
has smaller local fluctuations than $\rho^{\rm W}(Q)$ even though they are 
constructed using the same particles. 
Using a detailed error analysis it was determined that
pairs from the same W contribute a factor $1+c$ more to the 
final variance of the BEC measurements at low $Q$-values than
pairs from different W's. For this analysis a value of $c=1.9$
was determined and used in the following.

The contribution to the statistical variance was estimated for samples of
pairs with a given purity and was found
to be proportional to $1 + c \cdot (1-p(Q,\gamma,\theta^*))$.
Finally, all pairs were weighted by $p(Q,\gamma,\theta^*)$
divided by the above variance factor.
Using these weights, the improvement in the statistical error 
on the final measurement was 9\%. This is the reason for choosing the analysis 
which weights all particle pairs with their information content 
for the final result. The above procedure not only reduces the 
statistical error but makes the analysis more sensitive to 
pairs which originate from different W's. 

The two-particle densities in $Q$ for the combined data set are 
shown in figure~\ref{qplots} for both like-sign particle pairs and 
unlike-sign particle pairs, with and without pair weights. 
In both the fully-hadronic and semi-leptonic samples, the number of
unlike-sign pairs is higher than the number of like-sign pairs at $Q$ values
below 2 GeV. This is due to the large number of resonance decays 
with masses in this range. 
The region around $Q=0.7$ GeV is dominated by $\rm{\pi^+\pi^-}$ pairs
coming from the $\rho$ resonance which is abundantly present in hadronic
decays of the W. Reflections of three-body decays are also present in the
like-sign distributions. The two-particle densities of like-sign and
unlike-sign pairs for the mixed events
coincide, the reason for this being that all pairs in this
distribution contain particles from different events.
When pair weights are applied, contributions from resonance decays are down-weighted,
therefore the like-sign and unlike-sign spectra become more similar.

\subsection{Background subtraction}
The histograms in figure~\ref{qplots} show the contribution 
from $\rm{q\overline{q}(\gamma)}$ background events as they are simulated
with the BE$_{32}$ model.
The density $\rho^{\rm WW}(Q)$ is, consequently,
corrected for this background using the expression
\begin{equation}
\rho^{\rm WW}(Q) = \frac{1}{N_{\rm tot}-N_{\rm q\overline{q}}}\left(\frac{dn_{\rm tot}}{dQ}-\frac{
dn_{\rm q\overline{q}}}{dQ}\right),
\end{equation} 
\noindent
where $N_{\rm tot}$ and $N_{\rm q\overline{q}}$ are the total number of
selected events and the number of selected background events, 
respectively, and
$n_{\rm tot}$ and  $n_{\rm q\overline{q}}$ the respective number of particle
pairs from these  events.

The correlation strength parameter, PARJ(92), was re-tuned to a value of 
0.9 in order to get a better description of four-jet Z events, 
which are more similar to the selected background than inclusive Z decays.
This re-tuning was only used for the background events.
Details on the background tuning are discussed in section~\ref{sec:sec8}.
\subsection{Fit parametrisation}
\begin{table}[!t]
\begin{center}
\begin{tabular}{|c||c|c|c|c|c|} \hline
{\footnotesize sample/parameter}& 
{\footnotesize $\Lambda_{R_2}$}&
{\footnotesize $R(fm)$}&
{\footnotesize $\epsilon_d$}&
{\footnotesize $\delta_N$} &
{\footnotesize $\chi^2$/ndf}\\ \hline\hline
{\footnotesize BE32}& 
{\footnotesize 0.77 \p 0.02} &  
{\footnotesize 0.59 \p 0.01} & 
{\footnotesize -0.78 \p 0.02} &
{\footnotesize 0.013 \p 0.003} &  
{\footnotesize 141.8/96}\\\hline
{\footnotesize data}& 
{\footnotesize 0.64 \p 0.07} &  
{\footnotesize 0.59 } & 
{\footnotesize -0.78 } &
{\footnotesize 0.018 \p 0.017} &  
{\footnotesize 114.7/98}\\
\hline
\end{tabular}
\caption{Results of the fit, using equation~(\ref{laguerre}), to $R_2(Q)-1$,
  obtained from semi-leptonic WW decays, both for data and the BE$_{32}$ model.}
\label{r2results}
\end{center}
\end{table}

In order to quantify an excess at small $Q$ values, fits were 
performed to the inter-W correlation function $\delta_{\rm I}(Q)$. 
The choice of fitting function is motivated by the shape of 
the $R_2(Q)$ measurements. $\delta_I(Q)$ and $R_2(Q)$ have nearly 
the same physical meaning but different systematics since $R_2(Q)$
is dependent on the fragmentation model used.
This means that the optimal fitting functions are not necessarily 
the same and  that comparisons between $R_2(Q)$ and $\delta_I(Q)$ results must 
be done carefully. 

It is known from BEC measurements of the hadronic final state 
of a  Z or a single W 
that two particle correlation functions are reasonably well described by
either a Gaussian or an exponential parametrisation~\cite{LEPBECZ0,LEPBECWW}.
However, the BE$_{32}$ Monte Carlo samples show a more detailed 
structure in the $Q$ range between 0.5 and 1.5 GeV, as can be seen 
from figure~\ref{r2q}.
The slope observed around 1 GeV flattens out above
$Q=2$ GeV. Therefore in the plots of the correlation functions we restrict
to the $Q$ range 0-2 GeV. The fits are, however, performed in the
$Q$ range 0-4 GeV.
The dip around $Q=0.7$ GeV and the following slope of the correlation function
were treated as integral parts of the BE correlation function and as
integral parts of the BEC effect. 
Therefore, all $R_2(Q)$ distributions are fitted with the parametrisation

\begin{equation}
R_2(Q)-1=\Lambda_{R_2} e^{-RQ}(1+\epsilon_d R Q)+\delta_N, \label{laguerre}
\end{equation}
where $\Lambda_{R_2}$ denotes the correlation strength, $R$ is related to the
source size, and the term with $\epsilon_d$ accommodates the dip around 
$0.6 < Q < 1.0$ GeV.
The auxiliary term $\delta_N$ accounts for small differences in the charged
multiplicity of the signal and reference samples leading to a potential
bias in the normalisation.

The results of the correlation function, $R_2(Q)-1$, are summarised in
table~\ref{r2results} and shown in figure~\ref{r2q}. 
In order to compare the measured correlation strength,
$\Lambda_{R_2}$, between data and the BE$_{32}$ model, the $R$ and $\epsilon_d$
parameters were fixed to the values obtained from the model.
The measured correlation strength in data was found to be slightly smaller
than in the model. 
The significance of this difference is 1.8 standard deviations,
but it does not include systematic errors. Fitting measurements of 
$R_2(Q)$ must typically include an additional slope parameter and implies
that it 
takes into account the uncertainty due to fragmentation. 
This leads to a large correlation with the $\epsilon_d$ parameter and 
means that it is no longer possible to extract meaningful results 
on $\epsilon_d$, since it is not known whether to attribute the dip to 
the signal or the fragmentation of the reference sample. The $R$ parameter 
will also be affected by this but to a lesser degree and finally 
the $\Lambda_{R_2}$ parameter is the most stable. No reliable way has been 
found to quantify the systematic errors on the $R_2(Q)$ measurements,
and therefore only a qualitative agreement between BE$_{32}$ and the 
semi-leptonic data can be demonstrated. 
\begin{figure}[!t]
\hspace{-13. mm}
  \begin{tabular}{cc}
      \epsfxsize=8.5 cm 
\epsfbox{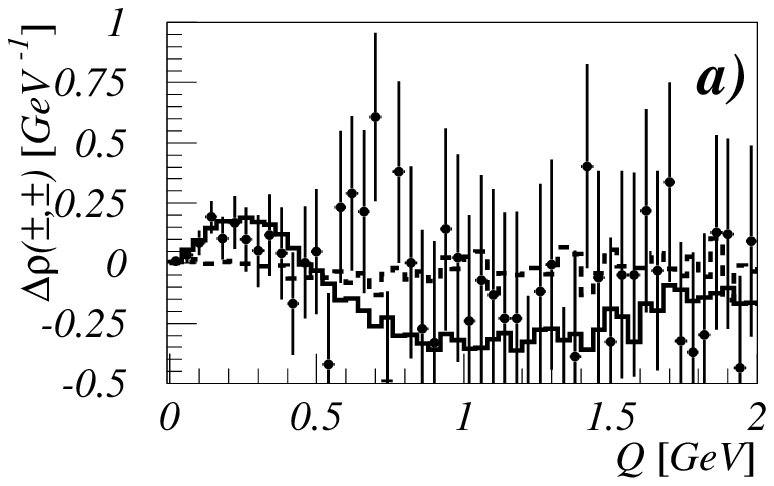}
&
      \epsfxsize=8.5 cm 
\epsfbox{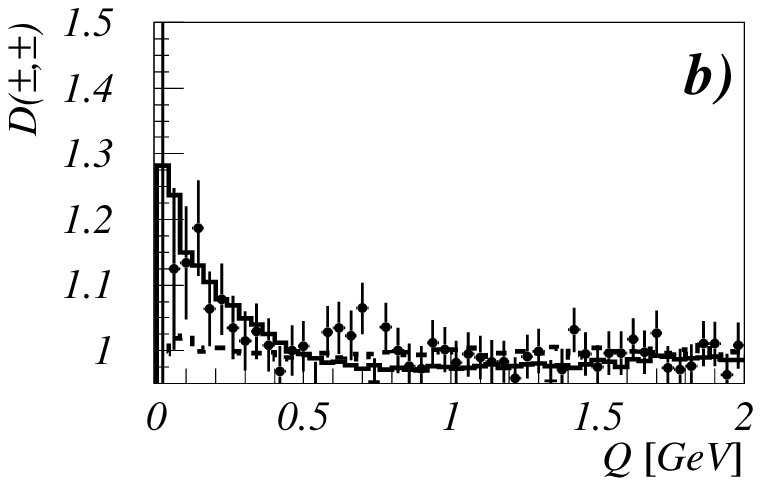}
\\
      \epsfxsize=8.5 cm 
 \epsfbox{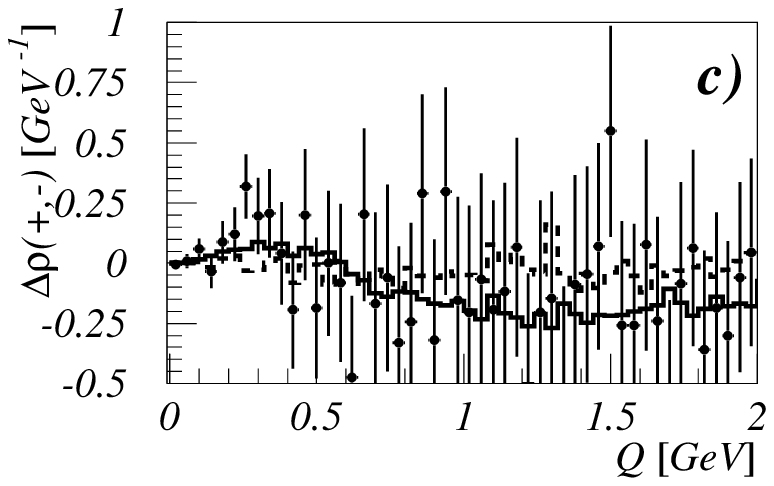}
      & 
      \epsfxsize=8.5 cm 
\epsfbox{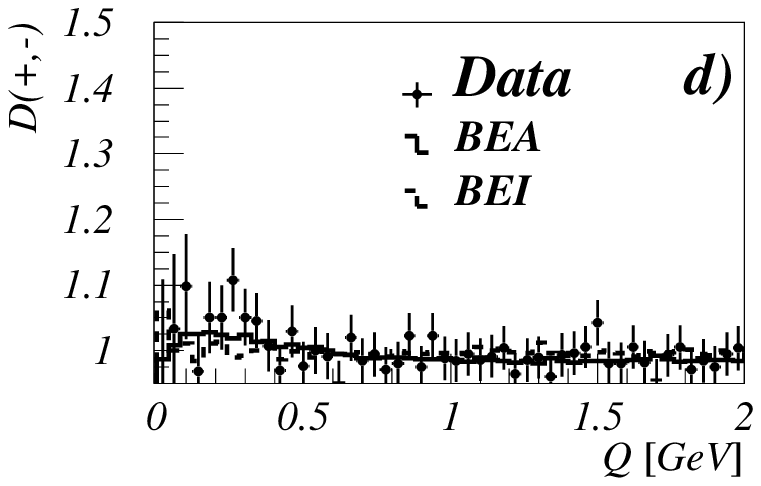}
  \end{tabular}
\caption{The $\Delta \rho (Q)$ (a,c) and $D(Q)$ (b,d) distributions for like-sign
  and unlike-sign particle pairs, respectively. The BE32 Monte Carlo models
  including BEC between all particle pairs (BEA), pairs coming from the
  same W (BEI) are superimposed on the plots.}
\label{delta}
\end{figure}

When fitting $\delta_I(Q)$ the normalization, $\delta_N$, can no longer be 
described as a simple additive parameter, and is included in the fits as:
\begin{equation}
\delta_I(Q) =  \Lambda_I e^{-RQ} (1+\epsilon_dRQ) +
\delta_N (1+\frac{\rho^{\rm W}(Q)}{\rho_{\rm mix}^{\rm WW}(Q)} ).
\label{newdifit}
\end{equation}

The comparison of $\delta_I(Q)$ and $R_2(Q)$ results is limited by
the systematics of the $R_2(Q)$ measurements, as described above.
Througout the paper, for reasons of clarity, the symbols $\Lambda_{R_2}$ and $\Lambda_{I}$ wil be
used when referring to fits of the $R_2(Q)$ and $\delta_I(Q)$ distributions respectively.
\subsection{Statistical errors and correlations}
\label{sect:sect6.3}
The values of the two-particle density distributions for different bins are
statistically correlated, since in general a particle occurs in several pairs and
because of non-Poissonian fluctuations
in the overall particle multiplicity~\cite{alessandro}.

The covariance matrix elements for the histogrammed spectra are given by
$V_{i,j} = \braket{h_i h_j}-\braket{h_i}\braket{h_j}$,
where $h_i$ and $h_j$ are the contents of bins $i$ and $j$.
These covariances are
propagated in the computation of the errors related to the distributions in 
equations~(\ref{drho})--(\ref{deli}). The use of pair weights  
does not pose any problems within this approach.
The statistical errors shown in the figures in this paper were computed only
from the diagonal elements of the covariance matrices.

Because of the limited precision of $V$, the covariance matrix was 
treated carefully from a numerical point of view in the subsequent fits.
The results and the covariance matrix were obtained from
the same data and therefore correlated. This correlation 
is effectively enhanced by the substantial correlations 
in $V$ and is a cause for biases in the fit results. 
By choosing suitable transformations of the fitting functions
this bias can often be reduced. However, for the $\delta_I(Q)$ observable
the value of $\Lambda$ is already nearly unbiased when no transformation is applied. 
All these biases were found to be inversely proportional to the number
of fitted events and the effect could therefore be estimated
by comparing the fit results on large samples to the fit 
results on data-sized samples. 
The final biases on the parameters $\Lambda$ and $\delta_N$ (see section
6.4) which were obtained for data-sized samples, 
were estimated to be $0.040+0.031\cdot\Lambda$ and $-$0.0121, respectively.
The biases on $\epsilon_d$ and $R$ were found to be 
completely negligible. 
All the final fit results were corrected for these biases.

The statistical errors were verified using resampling techniques
and were found
to be unbiased within a relative precision of 2\%.

\section{Results}
\label{sec:sec7}
\begin{figure}[!t]
\begin{center}
\begin{tabular}{cc} 
\hspace{-1.cm}
\mbox{\epsfig{file=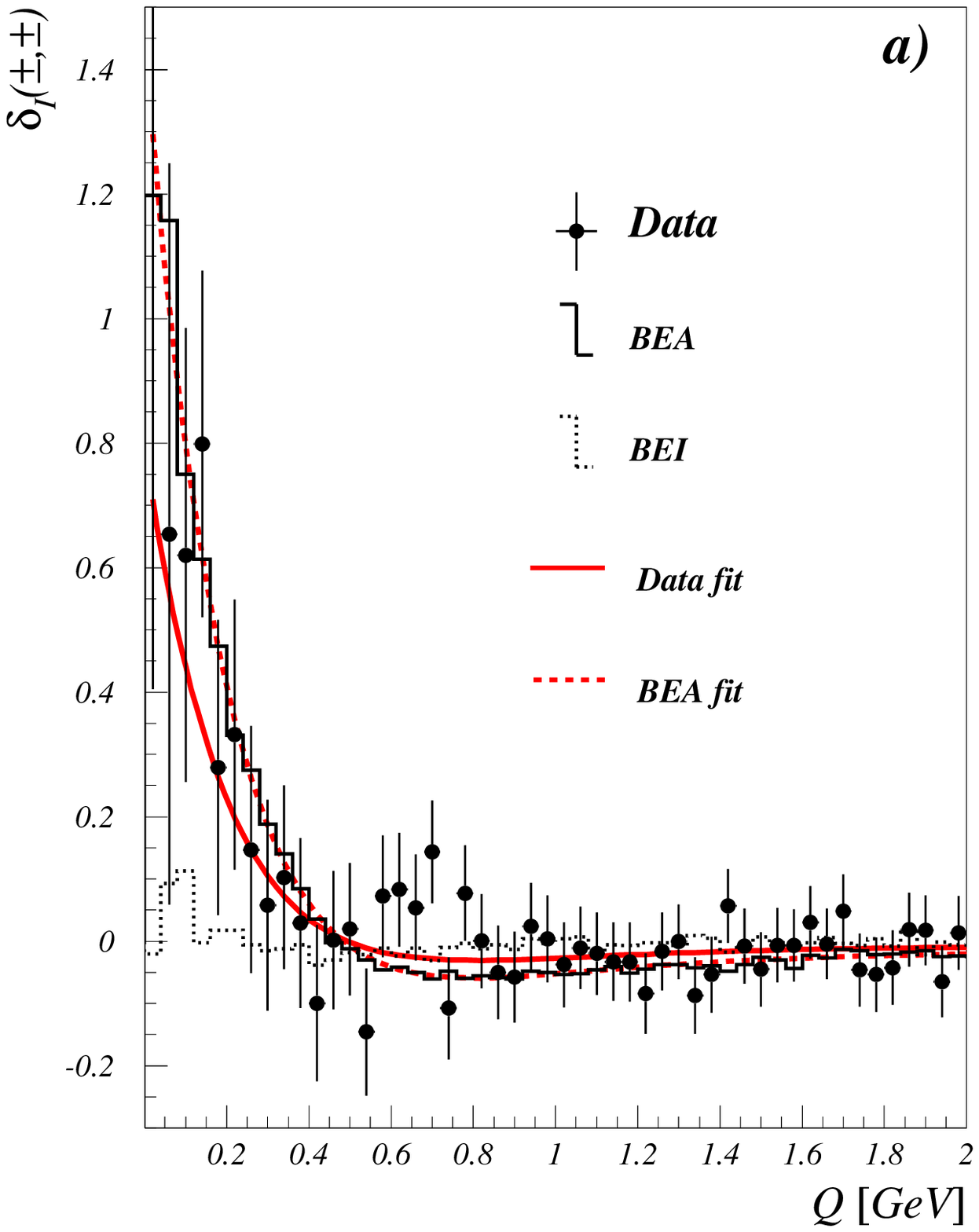,height=10.5cm}}
&
\hspace{-.5cm}
\mbox{\epsfig{file=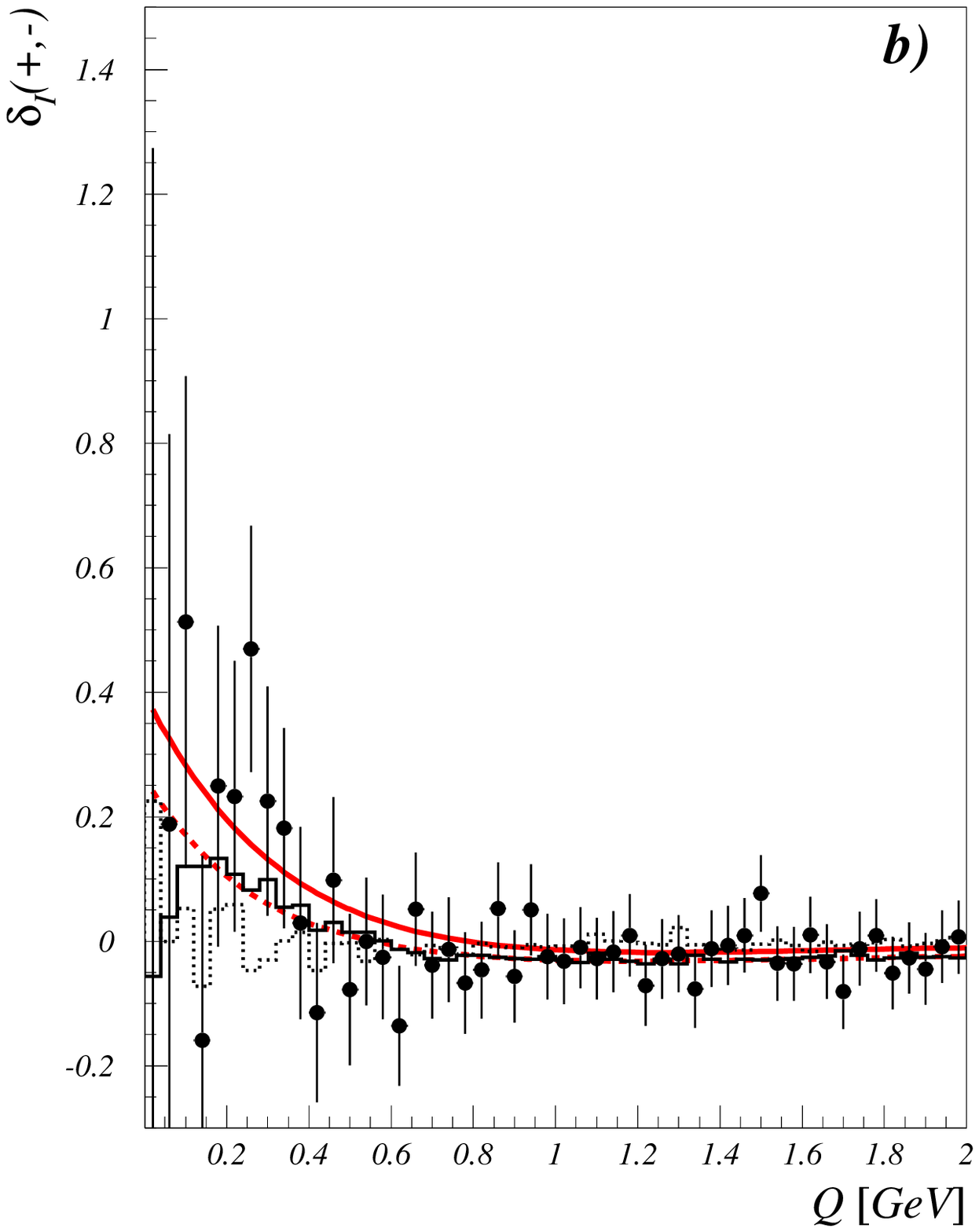,height=10.5cm}}\\
\end{tabular}
\caption{The $\delta_I(Q)$ distribution for like-sign (a) and unlike-sign (b)
particle pairs. The BE$_{32}$ models including BEC between all particle pairs
and pairs coming from the same W only are superimposed on the plots, together with the fit results
  using equation~(\ref{newdifit}). The fit results are with fixed $R$ and
  $\epsilon_d$ and are corrected for the bias mentioned in section~\ref{sect:sect6.3}.} 
\label{delifits}
\end{center}
\end{figure}
\begin{table}[!t]
\begin{center}
\begin{tabular}{|c||c|c|c|c|c|} \hline
{\footnotesize sample/parameter}& {\footnotesize $\Lambda_I$}&{\footnotesize
  $R(fm)$}&{\footnotesize $\epsilon_d$}&{\footnotesize $\delta_N$} &{\footnotesize
  $\chi^2$/ndf}\\ 
\hline\hline
\multicolumn{6}{|c|}{{\footnotesize $R$ free and $\epsilon_d$ fixed to BEA values}}\\\hline
{\footnotesize BEA (\p,\p)} & 
{\footnotesize 1.50 \p 0.06} &
{\footnotesize 0.72 \p 0.02} &
{\footnotesize -0.50 \p 0.03} &
{\footnotesize -0.010 \p 0.002} &
{\footnotesize 116.4/96} \\\hline
{\footnotesize Data (\p,\p)} & 
{\footnotesize 1.42 \p 0.63}&
{\footnotesize 1.14 \p 0.33}&
{\footnotesize -0.50} &
{\footnotesize -0.002 \p 0.020}&
{\footnotesize 88.3/97} \\\hline 
{\footnotesize BEA (+,--\xspace)} &
{\footnotesize 0.30 \p 0.03} &
{\footnotesize 0.41 \p 0.03} &
{\footnotesize -0.60 \p 0.09} &
{\footnotesize -0.010 \p 0.002} &
{\footnotesize 110.8/96} \\\hline
{\footnotesize Data (+,--\xspace)} &
{\footnotesize 0.43 \p 0.22} &
{\footnotesize 0.45 \p 0.15} &
{\footnotesize -0.60}&
{\footnotesize 0.000 \p 0.020}&
{\footnotesize 93.2/97}\\\hline
\multicolumn{6}{|c|}{{\footnotesize $R$ and $\epsilon_d$ fixed to BEA values}}\\\hline
{\footnotesize Data (\p,\p)} & 
{\footnotesize 0.82 \p 0.29}&
{\footnotesize 0.72}&
{\footnotesize -0.50}&
{\footnotesize  -0.005 \p 0.020}&
{\footnotesize 89.9/98}\\\hline
{\footnotesize BEI (\p,\p)} &
{\footnotesize 0.10 \p 0.05} &
{\footnotesize 0.72} &
{\footnotesize -0.50}&
{\footnotesize -0.009 \p 0.004}&
{\footnotesize 99.3/98}\\\hline
{\footnotesize BE0 (\p,\p)} &
{\footnotesize 0.02 \p 0.02} &
{\footnotesize 0.72} &
{\footnotesize -0.50}&
{\footnotesize -0.010 \p 0.002}&
{\footnotesize 98.0/98} \\\hline
{\footnotesize Data (+,--\xspace)} &
{\footnotesize 0.40 \p 0.18}&
{\footnotesize 0.41} &
{\footnotesize -0.60} &
{\footnotesize -0.001 \p 0.020}&
{\footnotesize 93.2/98}\\\hline
{\footnotesize BEI (+,--\xspace)} &
{\footnotesize 0.01 \p 0.03}&
{\footnotesize 0.41} &
{\footnotesize -0.60} &
{\footnotesize -0.005 \p 0.003}&
{\footnotesize 137.5/98} \\\hline
{\footnotesize BE0 (+,--\xspace)} &
{\footnotesize -0.04 \p 0.02}&
{\footnotesize 0.41} &
{\footnotesize -0.60} &
{\footnotesize -0.009 \p 0.002}&
{\footnotesize 138.0/98}\\\hline
\end{tabular}
\caption{Fit results to like-sign and unlike-sign $\delta_I(Q)$ with $R$ free and $R$
  fixed. Only statistical errors are shown.}
\label{results}
\end{center}
\end{table}

%
Inter-W BE correlations were studied as function of the observables
$\Delta \rho (Q)$, $D(Q)$ and $\delta_I(Q)$ as defined in equations~(\ref{drho})--(\ref{deli}). These results
are shown as function of $Q$ in figures~\ref{delta},\ref{delifits} and compared to predictions of
the LUBOEI model. 
In the like-sign distributions, an excess in data at low $Q$ values can be
observed. Numerical results were extracted from the $\delta_I(Q)$
distribution, shown in figure~\ref{delifits}. This choice was made since the 
$\delta_I(Q)$ is a genuine inter-W correlation function having a clear
physical interpretation. Other results are given for comparison.
Note that none of the results are corrected for pion purity or secondary
tracks, as was mentioned in section~\ref{sec:sec3}.

The $\delta_I(Q)$ distribution for like-sign pairs was fitted
using equation~(\ref{newdifit}).
The fits to the BEA model were performed with all four fit parameters left free.
The correlation strengths obtained from
the BEI and BE0 models were used to estimate a possible bias for the
measurement which is treated as systematic error (see section~\ref{sec:sec8}).
The result where both $\Lambda_I$ and $R$ were left free in the fit provides 
the most model-independent result. However, fixing the values of $R$ and $\epsilon_d$ to the ones predicted by LUBOEI
BE$_{32}$ tuned to DELPHI inclusive Z data, makes model comparisons easier
and contains the same statistical significance.  

The results on the data and models are summarised in table~\ref{results}.
The main result of this paper is then the observed 
correlation strength for like-sign pairs:
\begin{equation}
\Lambda_I (\pm,\pm) =0.82 \pm 0.29 {\rm (stat)} \pm 0.17 {\rm (syst)},
\end{equation}
with $R$ fixed to 0.72 fm. The first error is statistical and the second
error is systematic.
The evaluation of the systematic uncertainty of this result is discussed in section~\ref{sec:sec8}.
The expectation from the BEA model yields
\begin{equation}
\Lambda_{I_{BEA}} (\pm,\pm) =1.50 \pm 0.06 {\rm (stat)}.
\end{equation}
\begin{figure}[!t]
\hspace{-1.cm}
\mbox{\epsfig{file=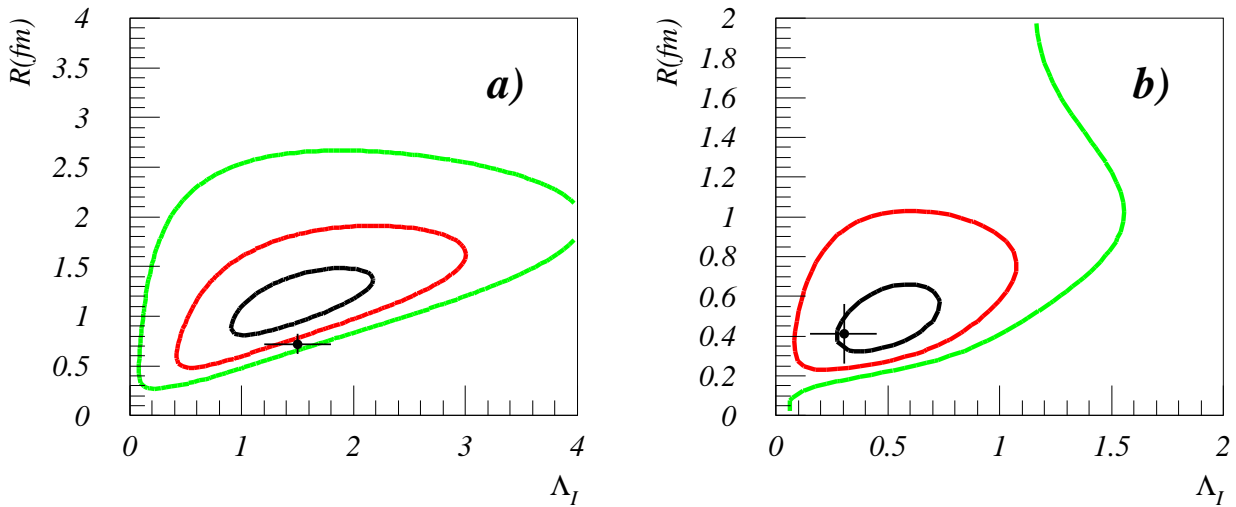,height=7.cm}}
\caption{The 2-d $\Delta \chi^2$ curves are shown for $\Lambda_I$ and $R$.
(a) shows the results for like-sign pairs, while
in (b) the unlike-sign pair result is displayed.
The three contours correspond to a $\Delta \chi^2$ of 
1, 4, and 9 respectively. The crosses show the prediction from LUBOEI 
BE$_{32}$. The errors corresponding to the LUBOEI predictions are multiplied
by a factor 5 for clarity.}
\label{fig:2d}
\end{figure}
The correlations in unlike-sign pairs were also measured
using the same procedure as for the like-sign pairs.
The result is shown in figure~\ref{delifits}~(b) and summarised in
table~\ref{results}. 
The distribution  shows some enhancements at low $Q$ when compared to the prediction from
BEI. Fitting with the same expression as for the like-sign pairs, but
with $R$ and $\epsilon$ fixed to the unlike-sign BEA prediction, yields:
\begin{equation}
\Lambda_I (+,-) = 0.40 \pm 0.18 {\rm (stat)} \pm 0.22 {\rm (syst)}
\end{equation}
in agreement with the expectation from BEA:
\begin{equation}
\Lambda_{I_{BEA}} (+,-) = 0.30 \pm 0.03{\rm (stat)}. 
\end{equation}
The LUBOEI model prediction for unlike-sign pairs has not been 
tuned on Z events as it was the case for the like-sign pairs.
Therefore this prediction should be treated more carefully
and an interpretation will be made in the discussion section.

The numerical results of the simultaneous fit of $R$ and $\Lambda_I$ are shown
in table~\ref{results}.
Since the two parameters are strongly correlated the results
are presented as $\Delta \chi^2$ curves in 2-d plots.
Figure~\ref{fig:2d} shows the fit results for both the like- 
and unlike-sign pairs. 
In these fits, the value of $\epsilon_d$ is still fixed since the data 
do not contain enough information about this parameter.
The position of the dip in the correlation function is hence
forced to scale like $R$. 
The systematic error is of the same size as the measurement with $R$ fixed
and not included in the fits shown in table~\ref{results} and figure~\ref{fig:2d}.
\section{Systematic uncertainties}
\label{sec:sec8}
The model-independent analysis of inter-W correlations
performed in this paper avoids
potential biases by the use of the event mixing and a
direct data-to-data comparison. Therefore only few sources of systematic
uncertainty remain to be considered; these are the subtraction of the $\rm{Z \rightarrow q\overline{q}}$
 background, the quality of the mixing of W final states
from
semi-leptonic events and the possible influence of colour reconnection.
\label{sect:bgsb}
\begin{figure}[!t]
\hspace{-13. mm}
  \begin{tabular}{cc}
      \epsfxsize=8.5 cm 
\epsfbox{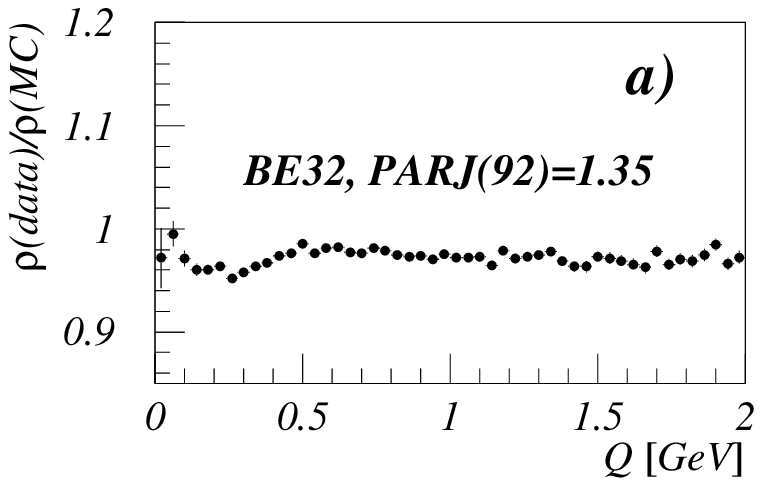}
&
      \epsfxsize=8.5 cm 
\epsfbox{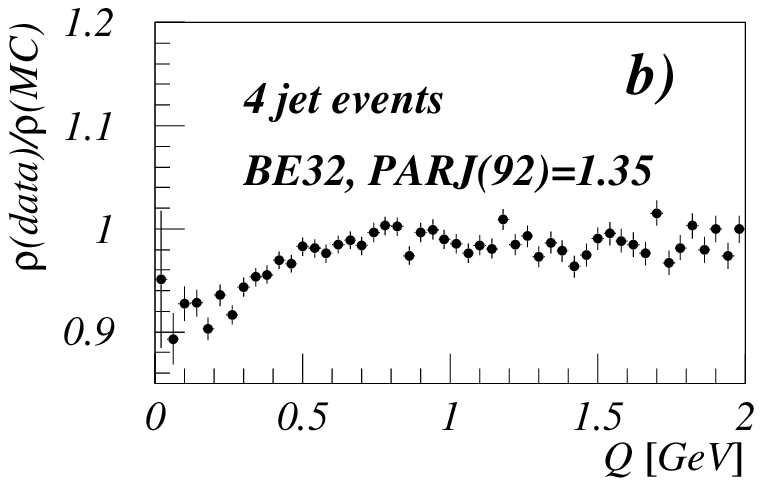}
\\
      \epsfxsize=8.5 cm 
\epsfbox{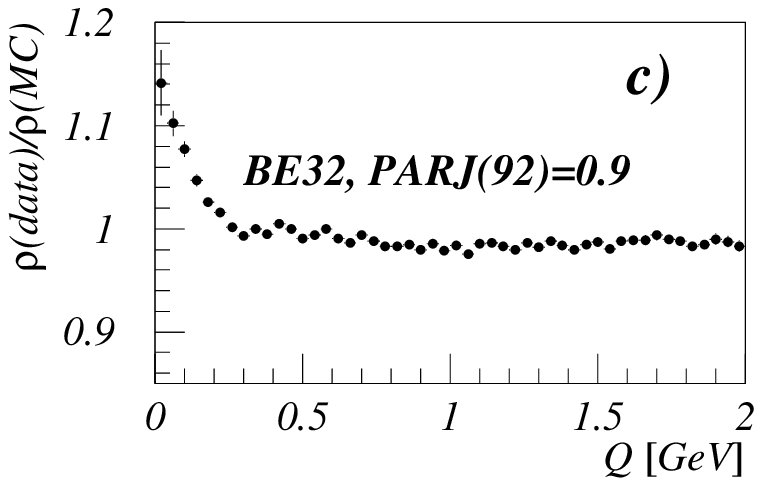}
      & 
      \epsfxsize=8.5 cm 
\epsfbox{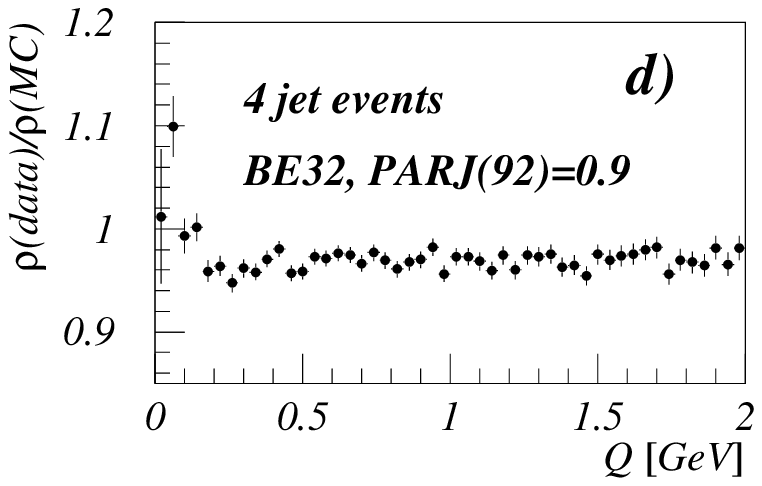}
  \end{tabular}
  \caption{Top plots: The ratio of the two-particle densities of
  like-sign particle pairs for Z events in data and in 
  Monte Carlo events using the BE$_{32}$ model  with parameters PARJ(92)=1.35 and
  PARJ(93)=0.34~GeV$^{-1}$, a) for the inclusive sample, b) for a four-jet sample with  
  $d_{34}>4.0$~GeV. The two bottom plots show the same comparison
  but with a different BE$_{32}$ input parameter PARJ(92)=0.9.}
  \label{bgtune} 
\end{figure}

  The background was subtracted using BE$_{32}$ simulation where the 
  correlation strength can be varied via the parameter PARJ(92).  
  Although the standard BE$_{32}$ tuning, using PARJ(92)=1.35, gives a good 
  description of inclusive
  Z-events, it was found that an input parameter strength of PARJ(92)=0.9
  gives a better description of Z-events having a clear four-jet
  topology. This can be seen in figure~\ref{bgtune} where both
  tunings are compared with inclusive Z-decays and four-jet like events.
  The four-jet events were selected with the
LUCLUS~\cite{Sjostrand:1995iq} clustering algorithm with
$d_{34}>4.0$~GeV.

  Both background samples were subtracted from the data and 
  half of the absolute difference in the final
  result of $\Lambda_I$, 0.075, was taken as a systematic uncertainty due to the lack of
  knowledge about BEC in $\rm{q\overline{q}(\gamma)}$ events.

In addition, data with a lower purity than at the working point were 
used to estimate the correlations in the background itself.
Four different purity bins were chosen and fits performed in each bin
to obtain the dependency of $\Lambda_I$ on the event purity.
The measured correlation strength depends non-linearly on the correlation
strength parameter PARJ(92) used in the LUBOEI simulation of the background.
A linear interpolation between the model predictions for the two
parameter choices of the following form has therefore been used:
Model($b$) = $b\cdot$ Model(PARJ(92)=1.35) + $(1-b)\cdot$
Model(PARJ(92)=0.90).
The extrapolation was found also to apply satisfactory 
outside the range $0<b<1$.

In each purity bin and for several different background subtractions
(as determined by different values of $b$)
the data were then fitted in order to obtain the corresponding
values of $\Lambda_I$ and $\partial\Lambda_I/\partial b$.
Only the uncorrelated errors were used in these fits, so that 
the results are independent except for the semi-leptonic 
events which are identical for each bin.
The results of these fits are shown in table~\ref{tab:back}.
The value of $b$ was then varied to obtain the background subtraction,
for which the $\Lambda_I$ values are independent of the purity.
The result of this variation gave the result: $b = -0.75 \pm 0.65$,
with $\chi^2=4.1$ for 3 degrees of freedom.

\begin{figure}[!t]
\hspace{-13. mm}
  \begin{tabular}{cc}
      \epsfxsize=8.5 cm 
\epsfbox{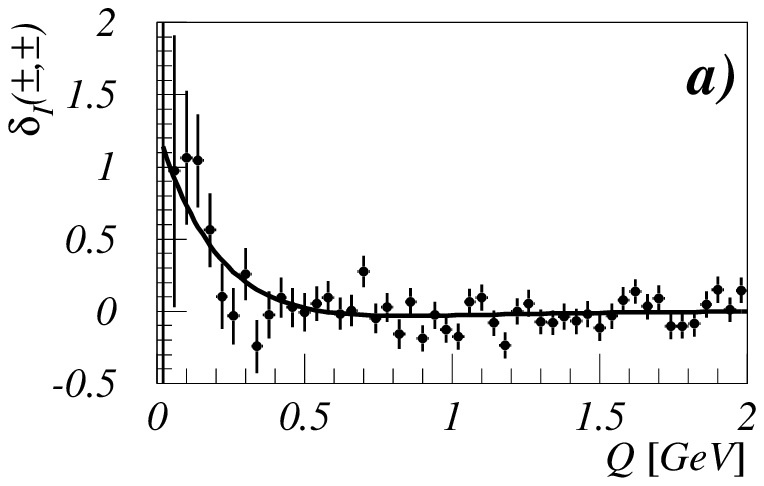}
&
      \epsfxsize=8.5 cm 
\epsfbox{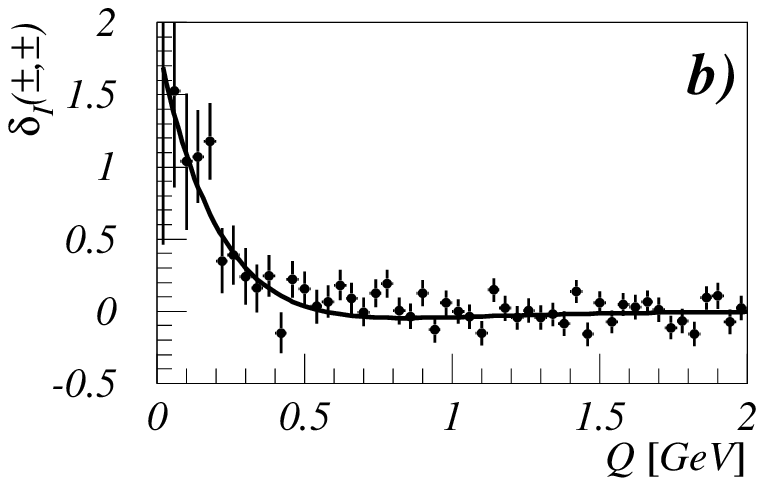}
\\
      \epsfxsize=8.5 cm 
\epsfbox{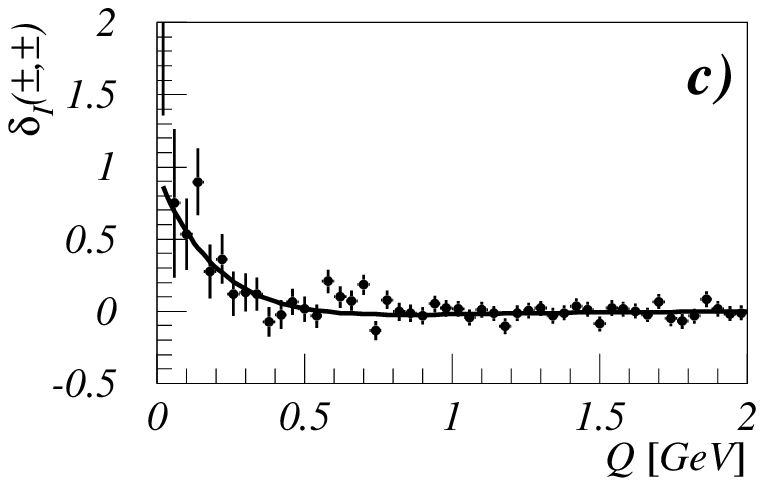}
      & 
      \epsfxsize=8.5 cm 
\epsfbox{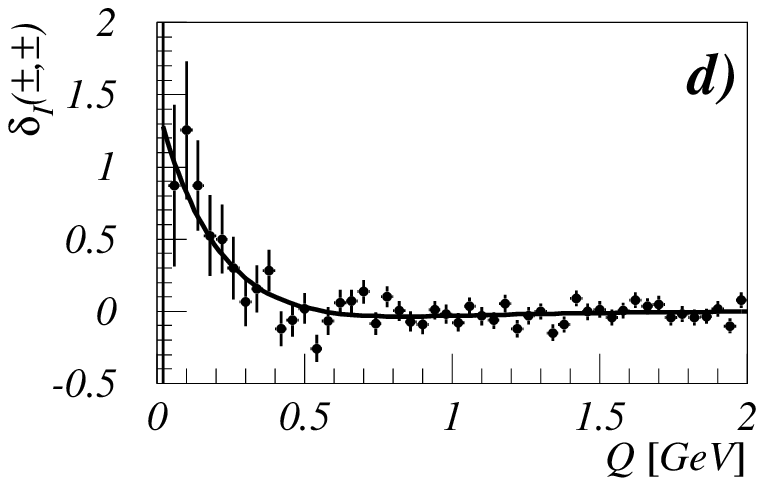}
  \end{tabular}
\caption{$\delta_{\rm I}$ is plotted for the 4 different purity bins of table~\ref{tab:back}
  (a-d).
The background is subtracted using $b=-0.75$. 
The results are strongly correlated between the purity bins.}
\label{fig:back}
\end{figure}

This value of $b$ is consistent with the value used, $b = 0.0 \pm 0.5$,
and is an additional strong indication that the correlations
in background 4-jet events are smaller than for the inclusive
sample ($b=+1$). The data are shown in figure~\ref{fig:back},
where the background with $b=-0.75$ is subtracted.
The amount of subtracted background was varied by 10\% (relative), which
changed $\Lambda_I$ by 0.007, which is negligible.

\begin{table}[!t]
\begin{center}
\begin{tabular}{|l|c|c|c|c|} \hline
Purity bin & a & b & c & d  \\ \hline 
Purity     & 0.60 & 0.77 & 0.89 & 0.97 \\
$\Lambda_I$ $(b=0)$ & 0.39 $\pm$ 0.36 & 1.28 $\pm$ 0.35 & 0.74 
$\pm$ 0.23 & 1.33 $\pm$ 0.33 \\
$\frac{\partial \Lambda_I}{\partial b}$ & -0.69 & -0.48 & -0.17 & -0.08 \\
$\Lambda_I$ $(b=-0.75)$ & 0.91 $\pm$ 0.40 & 1.64 $\pm$ 0.36 & 0.87 
$\pm$ 0.24 & 1.39 $\pm$ 0.33 \\ \hline
\end{tabular} 
\caption{Results for fits of individual bins in purity.
Only the uncorrelated errors are shown.}
\label{tab:back}
\end{center}
\end{table}

In the semi-leptonic events the background is a factor 2 smaller
than in the fully-hadronic events. 
It was verified that the topologies of these backgrounds are
identical to those of the signal, i.e.\ mostly 2-fermion decays into 
hadrons. These events behave with respect to BEC in the same way as inclusive
Z-events and therefore the BEC in these events are not expected to be 
significantly different than in the signal W-events. Finally, these 
background events do not suffer from the large extrapolation 
uncertainties of the 4-jet background in the fully-hadronic channel. 
The estimated systematic errors are shown in table~\ref{tabsyst}.

\label{sect:bgre}
  The selection of the data and the construction of the
  mixed reference sample may introduce distortions in the
  two-particle densities and therefore may lead to a non-zero value of
  $\Lambda_I$, measured in Monte Carlo samples without inter-W BEC.
  The fragmentation models used at LEP do not give a perfect 
  description of all the details of the soft fragmentation and 
  correlations. However, they constitute a reliable test for the
  magnitude of variation which can be expected for such effects.    
  The largest absolute value of the measured $\Lambda_I$, 0.125, for these
  models (PYTHIA~\cite{Sjostrand:1995iq}, HERWIG~\cite{HERWIG}, ARIADNE~\cite{ariadne}) was taken as a measure for the influence of selection
  procedures and mixing method on the measurement.
  The above method sums over many possible problems of the mixing.
In order to verify that the sum is not small due to an accidental
cancellation of large effects a weighting procedure was applied for 
several event variables and single particle distributions as described 
in section~\ref{sec:sec4}. 
BEI events were weighted so that the mixed and fully-hadronic events 
were in perfect agreement and the weighted events were refitted.
The maximum shift in $\Lambda_I$ was found to be 0.045 which is 
easily covered by the inclusive estimation.

Detector effects are very small due to the analysis method.
  Any simple variation of detector performance was found to give
  zero shift in $\Lambda_I$. Higher order effects are still possible,
  but these are also correlated with fragmentation properties
  and it was therefore assumed that they are covered
  by the previous estimate.       

A systematic uncertainty was attributed to the Colour
  Reconnection (CR) effect. This effect could have, in addition to
  inter-W BEC drastic consequences for the W mass
  measurement in the fully-hadronic channel~\cite{skmodel,arcr}.
  As for BEC, it violates the
  assumption that the two produced W bosons decay independently of each
  other.

  Colour reconnection occurs when independent colour singlets interact strongly
  before hadron formation.
  In fully-hadronic W decays it recombines partons from different parton
  showers. After fragmentation, the resulting hadrons carry therefore a 
  mixture of energy-momentum of both original W's.

  The CR effect has been modelled in various
  ways~\cite{skmodel,arcr,Lep2yellow}. Only the extreme 
  models~\cite{skmodel}, where  reconnection
  occurs in all events, have been ruled out by the LEP
  experiments~\cite{lepewnote}; however the absence of CR
  is also disfavoured by the same information.

  For this reason, three possible models of CR as
  implemented in JETSET, ARIADNE and HERWIG, were used to estimate their
  influence on this measurement. 
  The maximum difference in $\Lambda_I$ between the CR samples and 
  their equivalent models without CR implementation, 0.07, was found with the
  HERWIG implementation and was conservatively taken as systematic uncertainty due to
  the CR effect.
  
Finally, half of the bias correction (0.033) described in section~\ref{sect:sect6.3} 
was conservatively attributed as a systematic error due to fit biases.

The total systematic uncertainty on the measured $\Lambda_I$ value is
the sum in quadrature of the contributions listed in
table~\ref{tabsyst} for both like-sign and unlike-sign particle pairs.
\begin{table}[!t]
\begin{center}
\begin{tabular}{|c|c|c|} \hline
syst source & contribution to $\Lambda_I (\pm,\pm)$ & contribution to $\Lambda_I (+,-)$\\
\hline\hline
background BE model & 0.07 & 0.02\\\hline
semi-leptonic Bg. & 0.01 & 0.01 \\\hline
cuts \& mixing & 0.12 & 0.04\\\hline
Colour Reconnection  & 0.07 & 0.22\\\hline
Bias Correction & 0.03 &  0.03 \\\hline
Total syst. & 0.17 & 0.22\\\hline\hline 
\end{tabular}
\caption{A breakdown of the systematic errors for the $\Lambda_I$
  measurement with fixed $R$, for like-sign and unlike-sign particle pairs.}
\label{tabsyst}
\end{center}
\end{table}
\section{Discussion}
\label{sec:sec9}
The result, $\Lambda_I=0.82\pm 0.29\pm 0.17$, presented in section~\ref{sec:sec7} 
shows an indication for BEC in like-sign pairs between two hadronically 
decaying W bosons at the level of 2.4 $\sigma$ (standard deviations). 
The effect is 2.2 $\sigma$ below the prediction of BE$_{32}$, assuming that
the systematic uncertainties due to cuts and mixing are 100\% correlated. The 
spatial scale, $R$, of the correlations is higher at the level of 1.3 $\sigma$,
based on statistical uncertainties only. 
The QCD background to the WW signal is by itself of interest to study, 
and the data show that the BEC are weaker in these events than in 
inclusive hadronic Z events.

The measurement where both $R$ and $\Lambda_I$ are left free
and measured from the data contains the most model-independent 
information, while the result with fixed $R$ is optimal for 
testing the excess in the direction predicted by BE$_{32}$.

The results for unlike-sign pairs are difficult to interpret
due to a large model dependency and systematic errors.
The unlike-sign pairs show a smaller excess at low Q values. The excess
is at the 1.4 $\sigma$ level and in some agreement with the BE$_{32}$
prediction, but a bit too large if the unlike-sign effect is scaled
to follow the like-sign. 
The unlike-sign pairs are statistically 
correlated with the like-sign pairs at the level of 60\%.
Therefore it is not possible to rule out a large statistical 
component of these effects. 

The tuning of the BE$_{32}$ model was verified using 
the semi-leptonic WW events. 
The overall agreement including the 
description of the dip around $Q=0.7$ GeV was found to be excellent, 
except that the data
show a bit less correlation than the prediction tuned on Z's.
To this 1.8 $\sigma$ statistical difference a systematic
uncertainty which comes from extrapolating from Z
data using the model has to be added. This uncertainty which is 
difficult to control is also the reason why the main result of this 
paper does not rely on BE$_{32}$ simulation.

Inside the LUBOEI models the input parameter PARJ(92) determines the 
strength of the correlations. However, due to non-linear effects 
in the implementation, there is no one-to-one correspondence 
between the generated strength and the measured strength, $\Lambda_{R_2, I}$.
Even with a fixed value of PARJ(92) the observed $\Lambda_{R_2}$ does 
depend on the multiplicity of the event. The used value of
PARJ(92)=1.35 leads to a measured value of $\Lambda_{R_2}=0.77 \pm 0.02$
for the semi-leptonic BE$_{32}$ events while the BEA result is $\Lambda_I=1.50 \pm 0.06$.
The two BE$_{32}$ results come from fits to $R_2$ and $\delta_I$
respectively. The systematics are expected to be small on these 
numbers because they compare events which use the same fragmentation.
The above observation can be combined with the information that the 
improvement in the statistical sensitivity to BE$_{32}$ was smaller
(9\%) than the improvement (17\%) expected from pure statistics.
This indicates that the BEA not only adds correlations between
particles from different W's compared to BEI, but increases
the strength of all correlations significantly. This enhancement
scenario is not supported by the 4-jet Z-data where 
the BE$_{32}$ model clearly overestimates the correlations 
when tuned on inclusive events.

Both $\Lambda_{R_2, I}$ and $R$ are subject to additional 
corrections if their values are to be combined with other experiments. 
$\Lambda_{R_2, I}$ is diluted due to pair impurities to a level of
about 70\%, while $R$ is only comparable to either
other data or models. 

%
%
%
In conclusion this paper presents a model-independent measurement
of inter-W BEC.
 The measurement does assume that intra-W BEC are 
the same in 2-jet and 4-jet WW decays. 
The direction in which the signal is looked for
is motivated by the measured BEC in semi-leptonic WW events and
implemented via the BE$_{32}$ model tuned to Z data. 
The measurement is firmly based on the mixing of
the
hadronic final state of independent W bosons from semi-leptonic WW
events.
This technique yields small systematic uncertainties in spite of the
small fraction of particle pairs from different W's. A weighting
technique
allowed the effective purity to be raised to about 20\%.

The treatment of statistical errors and bin correlations of
the
correlation function allows for a reliable specification of the
statistical error of the quoted correlation strength. The remaining
systematic uncertainties and their dependence on model parameters
were carefully studied.
\section{Conclusion}
\label{sec:sec10}
Overall, there is an indication of correlations in like-sign pairs 
from different W's with a significance of 2.4 standard deviations.
The results are 2.2 standard deviations lower than the expectation 
of the BE$_{32}$ model which was tuned to DELPHI Z data and verified 
on semi-leptonic WW events. 
The spatial scale of the correlations is larger but consistent
with BE$_{32}$. The results from the unlike-sign pairs are inconclusive,
and their interpretation is model dependent.

The prediction of the LUBOEI model is disfavoured not only 
by the behaviour of the WW data but also through the 
investigations of the background. It is clear 
that developments in the theoretical side of fragmentation 
models are needed before the BEC results presented in this 
paper can be fully understood.

These final results of the DELPHI experiment have a limited
statistical
accuracy. The precision would be significantly increased by a combination with the 
results of the LEP experiments which can be found in~\cite{LEPBECWW}.
\subsection*{Acknowledgements}
\vskip 3 mm
 We are greatly indebted to our technical 
collaborators, to the members of the CERN-SL Division for the excellent 
performance of the LEP collider, and to the funding agencies for their
support in building and operating the DELPHI detector.\\
We acknowledge in particular the support of \\
Austrian Federal Ministry of Education, Science and Culture,
GZ 616.364/2-III/2a/98, \\
FNRS--FWO, Flanders Institute to encourage scientific and technological 
research in the industry (IWT), Belgium,  \\
FINEP, CNPq, CAPES, FUJB and FAPERJ, Brazil, \\
Czech Ministry of Industry and Trade, GA CR 202/99/1362,\\
Commission of the European Communities (DG XII), \\
Direction des Sciences de la Mati$\grave{\mbox{\rm e}}$re, CEA, France, \\
Bundesministerium f$\ddot{\mbox{\rm u}}$r Bildung, Wissenschaft, Forschung 
und Technologie, Germany,\\
General Secretariat for Research and Technology, Greece, \\
National Science Foundation (NWO) and Foundation for Research on Matter (FOM),
The Netherlands, \\
Norwegian Research Council,  \\
State Committee for Scientific Research, Poland, SPUB-M/CERN/PO3/DZ296/2000,
SPUB-M/CERN/PO3/DZ297/2000, 2P03B 104 19 and 2P03B 69 23(2002-2004)\\
FCT - Funda\c{c}\~ao para a Ci\^encia e Tecnologia, Portugal, \\
Vedecka grantova agentura MS SR, Slovakia, Nr. 95/5195/134, \\
Ministry of Science and Technology of the Republic of Slovenia, \\
CICYT, Spain, AEN99-0950 and AEN99-0761,  \\
The Swedish Research Council,      \\
Particle Physics and Astronomy Research Council, UK, \\
Department of Energy, USA, DE-FG02-01ER41155. \\
EEC RTN contract HPRN-CT-00292-2002. \\

\bibliography{newbib}
\end{document}